\documentclass[fleqn,usenatbib]{mnras}

\usepackage[T1]{fontenc}

\DeclareRobustCommand{\VAN}[3]{#2}
\let\VANthebibliography\thebibliography
\def\thebibliography{\DeclareRobustCommand{\VAN}[3]{##3}\VANthebibliography}

\usepackage{graphicx}	
\usepackage{newtxtext,newtxmath}
\usepackage{orcidlink}

\title[Failed winds and the BLR]{A dust condensation instability in AGN atmospheres: failed winds and the broad line region}

\author[Owen, J. E. \& Lin, D. N. C.]{
James E. Owen$^{\orcidlink{0000-0002-4856-7837}1}$\thanks{E-mail: james.owen@imperial.ac.uk} and Douglas N. C. Lin$^{\orcidlink{0000-0001-5466-4628}2, 3}$\\
$^{1}$Imperial Astrophysics, Department of Physics, Imperial College London, Blackett Laboratory, Prince Consort Road, London SW7 2AZ, UK\\
$^{2}$Department of Astronomy and Astrophysics, University of California, Santa Cruz, CA 95064, USA\\
$^{3}$Institute for Advanced Studies, Tsinghua University, Beijing, 100086, China
}

\date{Accepted XXX. Received YYY; in original form ZZZ}

\pubyear{2015}

\begin{document}
\label{firstpage}
\pagerange{\pageref{firstpage}--\pageref{lastpage}}
\maketitle

\begin{abstract}
Active galactic nuclei (AGN) are important drivers of galactic evolution; however, the underlying physical processes governing their properties remain uncertain. In particular, the specific cause for the generation of the broad-line region is unclear. There is a region where the underlying accretion disc atmosphere becomes cool enough for dust condensation. Using models of the disc's vertical structure, accounting for dust condensation and irradiation from the central source, we show that their upper atmospheres become extended, dusty, and radiation-pressure-supported. Due to the density--temperature dependence of dust condensation, this extended atmosphere forms as the dust abundance slowly increases with height, resulting in density and temperature scale heights considerably larger than the gas pressure scale height. We show that such an atmospheric structure is linearly unstable. An increase in the gas density raises the dust sublimation temperature, leading to an increased dust abundance, a higher opacity, and hence a net vertical acceleration. Using localised 2D hydrodynamic simulations, we demonstrate the existence of our linear instability. In the non-linear state, the disc atmosphere evolves into ``fountains'' of dusty material that are vertically launched by radiation pressure before being exposed to radiation from the central source, which sublimates the dust and shuts off the radiative acceleration. These dust-free clumps then evolve ballistically, continuing upward before falling back towards the disc under gravity. This clumpy ionized region has velocity dispersions $\gtrsim 1000$~km~s$^{-1}$. This instability and our simulations are representative of the Failed Radiatively Accelerated Dusty Outflow (FRADO) model proposed for the AGN broad-line region.
\end{abstract}

\begin{keywords}
accretion, accretion discs -- instabilities -- radiation: dynamics -- galaxies: active
\end{keywords}



\section{Introduction}
The accretion discs that make up active galactic nuclei (AGN) are some of the most luminous objects in the universe. These discs, which feed the supermassive black holes at the centres of galaxies, span a wide range of radial scales, from small sub-AU scales near the black hole's event horizon, to the large dusty tori present on sub-parsec to parsec length scales \citep[e.g.][]{Antonucci1993,Netzer2015,Padovani2017}. Commonly associated with the accretion disc is line emission from the broad-line region and the narrow-line region. These emission lines are representative of high density ionized gas \citep[e.g.][]{CS1988,Netzer1990}.  These emission lines are thought to be Doppler broadened, with the $\gtrsim 1000~$km~s$^{-1}$ of the broad emission lines indicating they are coming from closer in, from distances of order $\sim 10^{-2}-10^{-1}$~pc \citep[e.g.][]{Peterson2004,Peterson2006}. However, despite the observational progress in studying AGN and the broad-line region, there is no clear consensus on the physical mechanisms underlying the broad-line region \citep[e.g.][]{Sulentic2000,Kaspi2005}. We know that this region is close to the accretion disc that powers the AGN \citep[e.g.][]{GravityAGN2018,GravityAGN2021}, and its velocity structure implies that it is dominated by Keplerian rotation \citep[e.g.][]{Grier2013}; however, the lines are sufficiently broadened so that the expected double-peaked structure is typically absent \citep[e.g.][]{Done1996,Kollatschny2003}. Furthermore, the high densities suggested from the ionized emission ($\sim 10^{10}-10^{12}$~cm$^{-3}$, e.g. \citealt{Adhikari2016, Panda2018}) and covering fraction of order $\sim 0.3$ \citep[e.g.][]{Korista1997,Maiolino2001} has given rise to the concept that the broad-line region is an inhomogeneous, clumpy region containing ionized ``clouds'' that give rise to the line emission. The inferred density, along with the measured column depths \citep[e.g.][]{Ruff2012}, implies that the clouds are extremely small with a typical size of $10^{12}-10^{13}$~cm \citep[e.g.][]{Maiolino2010,Naddaf2021}.  

The outer edge of the broad line region is restricted by the location of the dusty torus \citep[e.g.][]{Netzer1993}; where the dust is sufficiently cool to be able to survive the intense irradiation from the central source. The broad lines are typically divided into two categories, high-ionization lines and low-ionization lines. Reverberation mapping of the low-ionization lines (such as H$\beta$ and MgII) has shown that they come from distances closer to the central black hole than the dusty torus \citep[e.g.][]{Suganuma2006}. Several mechanisms have been proposed to explain the origin of the broad-line region, including magnetically \citep{Blandford1982} or thermally driven \citep{Begelman1983} winds. Line-driven winds were proposed by \citet{Murray1995}, and this remains a likely candidate for the high ionization region. Inflow models have been proposed \citep[e.g.][]{Gaskell2016, Wang2017}, or instabilities in the accretion discs themselves \citep[e.g.][]{Collin2008,Wang2012}. 

An intriguing suggestion for the origin of the low-ionization broad lines has been proposed: a dusty ``failed'' radiation pressure driven wind. \citet{Czerny2011}, argued that the measured location of the low-ionization broad-line region using reverberation mapping happened to occur at an approximately fixed effective temperature of the accretion disc. This temperature of $1000-1500$~K, which would be representative of the temperature in the disc's atmosphere, is exactly the temperature at which dust is expected to condense \citep[e.g.][]{Elvis2002}. Since the opacity to optical/UV emission from the central source is much higher than that to the disc's outgoing thermal irradiation, the atmosphere develops a region that is optically thin to its own thermal emission but optically thick to the central source's radiation. This shielding allows the atmosphere to remain near the effective temperature, providing a spatial location where dust can condense. \citet{Czerny2011} and \citet{Baskin2018} have shown that in this region a dramatic increase in opacity due to dust condensation would lead the atmosphere to receive a sufficiently high acceleration to unbind it. However, as the disc atmosphere is accelerated to higher altitudes above the disc, it can be exposed to the central source, sublimating the dust, removing the radiative acceleration. In the ‘failed’ wind scenario \citep{Czerny2011}, clouds launched from the dust-condensing surface of the disc are accelerated approximately vertically. However, once exposed to the central source, they lose radiative acceleration, fail to reach escape velocity, and fall back toward the disc. This idea has been successful in explaining several observed correlations \citep{Czerny2016,Naddaf2020,Wu2025}. Phenomenological models of ballistic clouds launched from the accretion disc have successfully reproduced observed line profiles and covering fractions \citep{Czerny2015,Naddaf2020,Naddaf2021,Naddaf2022,NC2024,Naddaf2024}. Furthermore, the rapid increase in height of the dusty disc's surface expected from this scenario \citep{Baskin2018} appears to be consistent with recent interpretation of reverberation mapping observations, which appear to indicate a strongly flared structure to the disc's surface at the location of the low-ionization broad-line region \citep{Starkey2023}. 

Although the high density of the broad-line region clouds can be explained by compression of an existing clumpy medium by radiative acceleration from the central source \citep{Baskin2014,Baskin2018}; what initially leads to the launch of a clumpy, radiatively driven dusty wind remains unexplored. Furthermore, previous modelling approaches have assumed that dust condensation or sublimation is essentially spatially or temporally instantaneous. Although the timescale for dust sublimation, when exposed to the central source, is rapid \citep{Baskin2018}; the density dependence of the dust sublimation temperature means that this is not guaranteed in the shielded regions of the disc's atmosphere. Thus, given the appealing nature of the dusty failed-wind scenario, here we explore the underlying physical mechanism that leads to the launch of a clumpy dusty wind from an AGN disc's atmosphere. We demonstrate that the density dependence of dust condensation leads to a dramatically inflated disc atmosphere where dust condensation slowly increases with altitude. Furthermore, again arising from the density dependence of dust condensation, we show that AGN atmospheres are linearly unstable to a compressible instability, naturally giving rise to a clumpy dusty wind launched from the disc's atmosphere. 

In Section~\ref{sec:smooth} we outline the basic physics of how density- and temperature-dependent dust condensation leads to dust condensation being spatially extended rather than occurring at a fixed altitude. In Section~\ref{sec:disc_vert} we construct numerical models for the vertical structure of the disc and analyse their linear stability in Section~\ref{sec:instability}. Section~\ref{sec:simulations} shows preliminary hydrodynamic simulations of dusty AGN atmospheres. We discuss the implications of our results in Section~\ref{sec:discuss}, before summarising in Section~\ref{sec:summary}.

\section{Dust condensation in an AGN disc atmosphere}\label{sec:smooth}
In this section, we begin our exploration of the implications of dust condensation in the upper layers of an AGN disc. Although dust condensation is often assumed to occur effectively instantaneously at a fixed temperature, in reality it occurs over a finite range of temperatures that vary (slowly) with pressure. The fact that dust condensation is a continuous process has important consequences, as we shall demonstrate with a schematic model (and later with full numerical models). Namely, that dust condensation does not occur at a fixed height in the disc but rather occurs continuously over a large vertical region (many pressure scale heights) in AGN atmospheres. To illustrate how smooth dust condensation affects the disc’s vertical structure, consider a simple 1D vertical atmosphere. Here the flux ($F$) is constant with height ($z$) and defines an effective temperature $T_{\rm eff}$. Due to the high opacity from dust and the high accretion luminosities, the atmosphere will be radiation pressure supported, where, for now we assume:
\begin{equation}
    \frac{\kappa\left(\rho,T\right)F}{c} \approx \Omega^2 z \label{eqn:force_balance_arad}
\end{equation}
where $\kappa$ is the density ($\rho$) and temperature ($T$) dependent opacity per unit mass of the dust and gas mixture, $\Omega$ is the disc's angular velocity (which we take to be vertically constant), and $c$ is the speed of light. Neglecting external irradiation, the temperature--optical depth relationship in the atmosphere obeys the approximate $T(\tau)$ relationship \citep{Hubeny1990}:
\begin{equation}
    T^4 = \frac{3}{4}T_{\rm eff}^4\left(\tau_z+\frac{1}{\sqrt{3}}\right) \label{eqn:T_tau}
\end{equation}
where $\tau_z$ is the optical depth measured from the top of the disc in the vertical direction ($\hat{\mathbf{z}}$). Now, for algebraic simplicity, assuming an opacity-law of the form:
\begin{equation}
    \kappa \propto \rho^{a}T^{b}, \label{eqn:opac_simple}
\end{equation}
the density gradient for a radiation-pressure supported disc becomes:
\begin{equation}
    \frac{{\rm d}\log \rho}{{\rm d}z} = \frac{3b}{16a} \frac{1}{\ell_{\rm mfp}} - \frac{1}{a z} \label{eqn:density_gradient}
\end{equation}
where $\ell_{\rm mfp}$ is the local photon mean-free-path, $1/(\kappa\rho)$.
\begin{figure}
    \centering
    \includegraphics[width=\columnwidth, trim=0.1cm 0 0 0, clip]{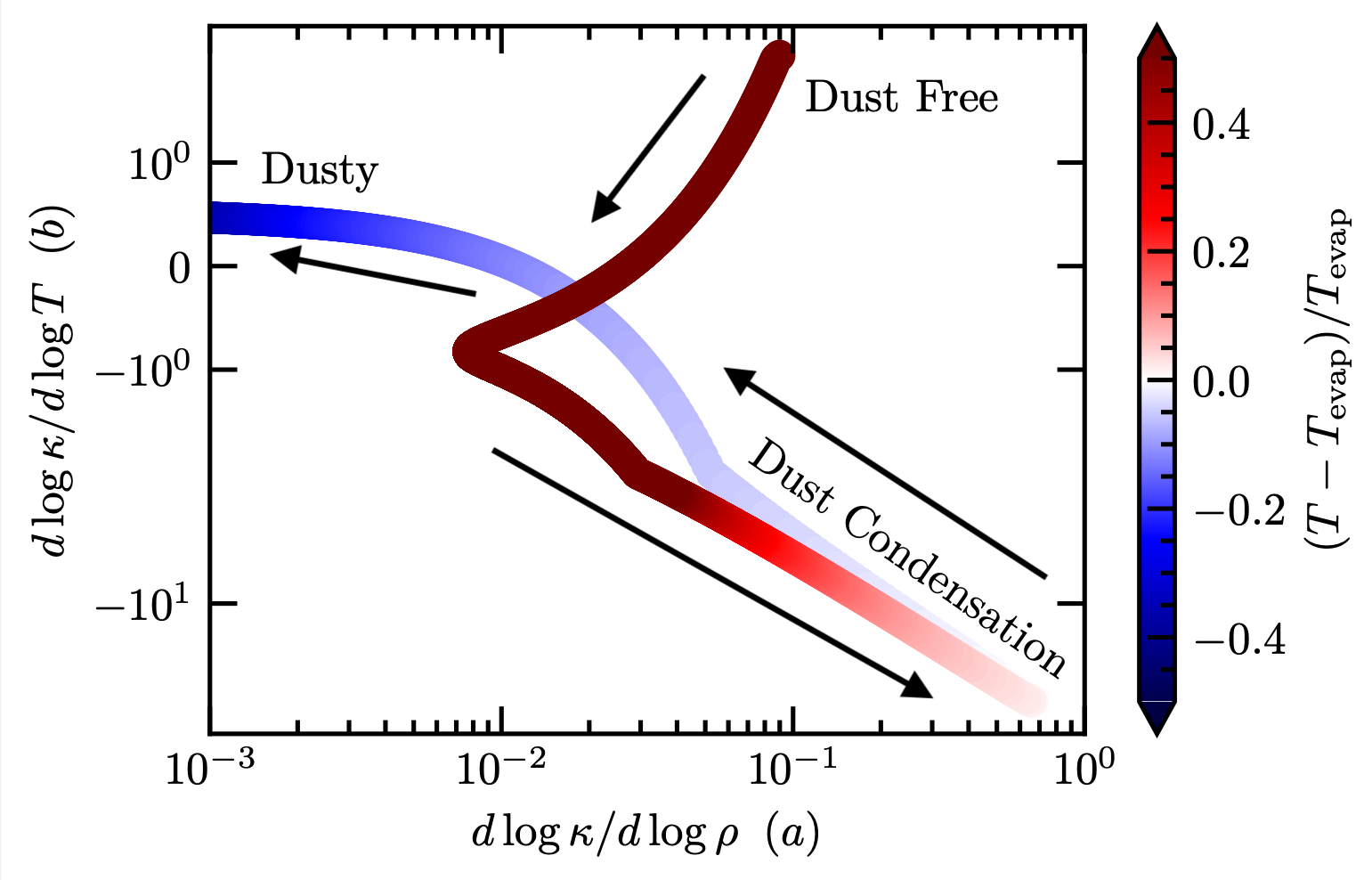}
    \caption{The variation of the density and temperature opacity indices ($a$, $b$ -- Equation~\ref{eqn:opac_simple}) during the dust condensation process at a gas density of $10^{-12}$~g~cm$^{-3}$, from dust free (top right) to dusty (top left),  as indicated by the arrows. This specific curve is plotted for our dust condensation model described in Section~\ref{sec:opacity} as the temperature is varied; however, whatever the specific model, the general result is that during dust condensation, $b$ becomes large and negative, and $a$ becomes order of unity.  }
    \label{fig:opac_index}
\end{figure}
Using the evolution of the opacity indices with temperature (Figure~\ref{fig:opac_index}) through the dust condensation process, Equation~\ref{eqn:density_gradient} implies that the density gradient can become shallow, having a length scale (${\rm d }z/{\rm d}\log \rho$) of order $z$ or larger, in the limit where the photon mean-free path becomes long and $a$ approaches unity. These conditions are satisfied during the dust condensation process, in the disc's optically thin atmosphere. If the density slowly varies with height, then so does the temperature, thus, to maintain force balance (Equation~\ref{eqn:force_balance_arad}), the opacity must slowly increase through dust condensation as the height increases. Equation~\ref{eqn:force_balance_arad} again implies that this must happen over a length scale of order $z$. Therefore, if the local photon mean-free path is comparable to or larger than the disc height, radiation pressure can support a vertically extended atmosphere. In this case, density and temperature decrease gradually with height, while dust abundance increases. In the vicinity of the disc's atmosphere, above the photosphere, the photon mean-free  path is becoming larger than the height of the disc. In a model that does not include the density dependence of the dust sublimation temperature ($a \rightarrow 0$), one also obtains a sharp transition as the length scale over which density (and thus optical depth) changes becomes very small since $b$ is always a large negative value during dust condensation. 

The above discussion implies that AGN atmospheres likely have extended dust condensation in a radiation-pressure-supported disc atmosphere. In this region, the density and temperature vary on length scales of order $z$ or larger. Compared to the gas-pressure scale height ($H$), which is typically $H/R\lesssim 0.01$ in AGN discs (\citealt{Goodman2003}, where $R$ is the radial distance to the central supermassive black hole), this dust condensation could occur over many 10s to 100s of pressure scale heights.   However, to demonstrate this extended atmosphere occurs in reality, and to study its properties in detail, we need numerical calculations of the disc's vertical structure.

\section{Numerical calculations of disc vertical structure}\label{sec:disc_vert}
In order to fully understand the vertical structure of the AGN disc in the region where its effective temperature is comparable to the dust condensation temperature, we must solve for its properties numerically. The disc is in a vertical dynamical balance between gravity, thermal gas pressure ($P$) and radiation pressure, so that:
\begin{equation}
    \frac{{\rm d}P}{{\rm d}z} = - \rho \Omega^2z +\rho \frac{\kappa F_z}{c}\label{eqn:pressure_full}
\end{equation}
where $F_z$ is the vertical flux. Since we restrict our calculations to vertical scales much smaller than $R$, we adopt the $z \ll R$ approximation for vertical gravity with $\Omega = GM_\bullet/R^3$ (where $M_\bullet$ is the mass of the supermassive black hole).  Since our focus is on the disc's atmosphere rather than the details of the disc properties near the midplane, we assume that all accretional heating is released at the mid-plane, where we inject a flux $F_{\rm acc}$ to represent the energy release from accretion. In this case, the frequency-integrated moments of the radiative transfer equation obey \citep[e.g.][]{Jankovic2021}:
\begin{eqnarray}
    \frac{{\rm d}J}{{\rm d}z} &=& - \frac{3\rho\kappa}{4\pi\lambda(J)}F_z\\
    \frac{{\rm d}F}{{\rm d}z} &=& \Gamma_{\rm irr}\\
    \frac{{\rm d}\tau_z}{{\rm d}z} &=& \rho \kappa
\end{eqnarray}
where $J$ is the mean intensity, $\Gamma_{\rm irr}$ is heating due to the central source (described in Section~\ref{sec:central}) and $\lambda(J)$ is our closure to the radiative transfer equations. Further adopting local energy balance such that:
\begin{equation}
    4\rho\kappa\left(\sigma_{\rm SB} T^4 - \pi J\right) = \Gamma_{\rm irr} \label{eqn:energy_balance}
\end{equation}
with $\sigma_{\rm SB}$ the Stefan-Boltzmann constant. We choose to adopt flux-limited diffusion with a flux limiter of the form:
\begin{equation}
    \lambda = \frac{2 + R}{6 + 3R +R^2}
\end{equation}
where $R$ is a measure of the photon mean-free path:
\begin{equation}
    R = \frac{\xi}{\rho\kappa}\frac{|{\rm d} J/{\rm d}z|}{J}
\end{equation}
The $\xi$ parameter is related to the Eddington Tensor. In the flux-limited diffusion problem, this is set to a constant. The standard choice is $\xi=1$, as it results in radiation propagating at the speed of light in the full-time-dependent radiative transfer problem in the free-streaming limit. However, the choice $\xi=1$ produces the incorrect temperature structure in plane-parallel atmospheres when in equilibrium \citep[e.g.][]{DobbsDixon2012}. Since we focus on equilibrium solutions and are not concerned with the time dependence of the radiation field, we instead choose $\xi=2$, which reproduces the correct temperature solution in a plane-parallel atmosphere \citep[e.g.][]{Schulik2023}. With this choice, we still reproduce the optically thick limit correctly and the temperature profile in the optically thin limit. However, our results are insensitive to the choice, and the same general profiles are obtained even if we assume radiative diffusion everywhere. 

\subsection{Opacity Model}\label{sec:opacity}
For the gas opacities, we adopt the analytic formalism of \citet{Bell1994}, where we introduce an additional factor, linearly scaling the gas opacities which are metal-dominated with the metallicity. However, we replace the dust opacity with a new formalism. Dust opacities are dependent on the composition and particle size distribution, which are complications which we neglect at this stage. Instead, we adopt the simple form:
\begin{equation}
    \kappa_d = 25 {\rm\, cm}^2{\rm \, g}^{-1} X_d \left(\frac{Z}{Z_\odot}\right)\left(\frac{T}{1000~K}\right)^{1/2}
\end{equation}
where $Z$ is the metallicity and $X_d$ is the fraction of condensed dust with $X_d=1$ fully condensed and $X_d=0$ fully sublimated. This form of the dust opacity gives the approximate power-law dependence seen in more complicated models at temperatures of a few hundred to thousand K \citep[e.g.][]{Bell1994}, and is slightly larger than typical silicate opacities to account for the presence of graphite grains \citep[e.g.][]{Baskin2018}.  
For the fraction of condensed dust, following \citet{Kuiper2010}, we adopt the following functional form:
\begin{equation}
    X_d = \frac{1}{2} + \frac{1}{\pi}\arctan\left(\frac{T-T_{\rm evap}(\rho)}{\Delta T}\right) \label{eqn:dust_condense_simple}
\end{equation}
where $T_{\rm evap}$ is the density dependent temperature at which the dust is 50\% condensed. Building a simple equilibrium condensation-sublimation model \citep[e.g.][]{Owen2020_snow}, and adopting the following vapour-pressure relationship of:
\begin{equation}
    P_{\rm vap}= \exp\left(-\frac{60000~{\rm K}}{T}+37.7\right) \, {\rm dyn\,cm}^{-2}
\end{equation}
which is an appropriate approximate relationship for a mixture of possible dust compositions \citep{Campos2024},
we find that $T_{\rm evap}$ is well fit by the following expression:
\begin{equation}
    T_{\rm evap} = 2150 ~{\rm K} \left [\left(\frac{\rho}{1~{\rm\, g}~{\rm \,cm}^{-3}}\right)+10^{-18}\right]^{0.0195}
\end{equation}
This expression is similar to others found in the literature \citep[e.g.][]{Isella2005}, where we have introduced the constant density factor of $10^{-18}$~g~cm$^{-3}$, since at low temperatures sublimation is energetically favoured but too slow to be relevant. We choose this density by comparing the sublimation timescale of a few micron-sized grains to a typical dynamical timescale. Strictly speaking, one should vary this density factor with the radius in the disc to account for the changing dynamical timescale; however, for simplicity, we choose a constant density. For our simple single composition equilibrium condensation-sublimation model, we find that $\Delta T\sim 15$~ K is a good match. However, to account for the fact that real astronomical dust has a range of sizes and compositions, we increase $\Delta T$ and set it to 25~K. Unsurprisingly, given the discussion of smooth dust condensation in Section~\ref{sec:smooth}, the width of the dust condensation does have a quantitative impact on our results (as it changes $b$); however, our qualitative conclusions remain unchanged and the basic physical processes are robust provided $\Delta T$ is larger than a few Kelvin. We note that Equation~\ref{eqn:dust_condense_simple} is a poor match to the condensation-sublimation model at low condensed fractions $\lesssim 10^{-3}$; however, we find in our calculations at these low condensed fractions that dust is effectively unimportant for the opacity structure at this point. 

\subsection{Irradiation by the central source}\label{sec:central}

The irradiation emitted from the central region of the AGN disc typically dominates the radiation locally emitted in regions where dust condensation is expected to occur. However, this radiation is unable to penetrate deeply towards the disc mid-plane, heating only the surface layers. This is similar to the irradiation of a passively heated, flared, circumstellar disc \citep[e.g.][]{Chiang1997}. Therefore, our irradiation model is motivated by approaches in circumstellar discs. Radiation from the central source strikes the upper layers of the disc at some grazing angle ($\phi$); absorbed radiation is then assumed to propagate vertically through the disc. Thus, the attenuation of this radiation can be approximated as $\tau_{\rm irr}=\gamma' \tau_z/\mu$, where $\gamma' = \kappa_{\rm irr}/\kappa$ is the ratio of the opacity to the irradiation ($\kappa_{\rm irr}$) to the opacity of the disc's local radiation field ($\kappa$) and $\mu = \sin\phi$. Thus, the heating rate due to the irradiation is locally given by:
\begin{equation}
    \Gamma_{\rm irr} = \gamma' \rho F_R \exp\left(-\tau_{\rm irr}\right)\label{eqn:irradiation_heating}
\end{equation}
where $F_R$ is the flux from the central source. In reality, $\phi$ depends on the global structure of the disc and must be calculated in a self-consistent manner via iteration \citep[e.g.][]{Chiang2001,Jankovic2022}. However, here, we treat it as a free parameter and explore its effect on the vertical structure. Finally, to further simplify the problem and remain agnostic to the exact nature of the radiation from the central source, we set $\gamma'$ to a constant\footnote{This makes the numerical solution of the problem easier by making the RHS of the governing ODEs fully analytic, resulting in more stable integrations.}, nominally choosing a value of 300, appropriate for UV irradiation interacting with material with a temperature of 1,000-2,000~K as appropriate in the region of interest \citep[e.g.][]{Ali2021}. This choice has a minor quantitative influence on our vertical disc structures.

\subsection{Solution}
Equations \ref{eqn:pressure_full}-\ref{eqn:energy_balance}, together with our opacity model, irradiation model, and the ideal gas law, form a closed system for a given vertical flux ($F_z(z=0)$) at the midplane. This mid-plane vertical flux typically represents the accretional heat released and is given by:
\begin{equation}
    F_z(z=0) = F_{\rm acc} =  \frac{3}{8\pi}\dot{M}\Omega^2
\end{equation}
with $\dot{M}$ the accretion rate. To solve for the vertical structure of the disc, we follow a similar idea to \citet{Jankovic2021} \& \citet{Jankovic2022}. It is easier to specify the disc's boundary conditions at its top rather than the mid-plane. We fix the pressure to $10^{-18}$ dyn cm$^{-2}$, following \citet{Jankovic2021}. The exact value is unimportant as long as it is sufficiently small. The flux is set to the sum of the flux released at the mid-plane and the absorbed (and therefore re-emitted) irradiation flux such that:
\begin{equation}
    F\left(z=z_{\rm top}\right)=F_z+\mu F_R
\end{equation}
The mean intensity is set to the plane parallel solution\footnote{Note here our previous discussion regarding the choice of Eddington Tensor in our flux-limited diffusion approach, where the standard choice would be inconsistent with this boundary condition.}:
\begin{equation}
    J(z=z_{\rm top}) = \frac{1}{2\pi} F\left(z=z_{\rm top}\right)
\end{equation}
and the vertical optical depth is set to 0. Given these boundary conditions, the disc structure ODEs can be vertically integrated to the mid-plane, given a value for the height of the disc's surface, $z_{\rm top}$. We do this using an RK45 integrator to a relative tolerance of $10^{-8}$ with a maximum step-size of 0.003 $z_{\rm top}$. The disc's surface is then solved for by requiring that either i) for an MRI $\alpha$ parameter of 0.01 the total dissipation through the disc matches the required value for the given accretion rate, or ii) if the condition for i) results in a gravitationally unstable disc, the value of $\alpha$, now assumed to rise from angular momentum transport due to self-gravity, is increased until the Toomre $Q$ parameter is equal to unity. Where we evaluate $Q$ as:
\begin{equation}
    Q = \frac{\tilde{c}_s(z=0)\Omega}{\pi G \Sigma}
\end{equation}
where $\tilde{c}_s(z=0)$ is the effective mid-plane sound-speed arising from thermal support and radiation pressure such that:
\begin{equation}
    \tilde{c}_s = \sqrt{\frac{k_bT}{\mu_g m_h}+\frac{4}{3}\frac{\sigma T^4}{c\rho}}
\end{equation}
with $k_b$ the Boltzmann constant, $\mu_g$ the gas' mean molecular weight and $m_h$ the mass of hydrogen. Finally, iii) if the required $\alpha$ to reach $Q=1$ exceeds the fragmentation threshold of $0.3$ \citep[e.g.][]{Goodman2003}, we introduce an extra vertical flux in excess of the viscous flux to mimic the production from other sources (e.g. star formation, \citealt{Shlosman1989,Goodman2003}). We find that the vast majority of the models we discuss in the following fall into categories (i) and (ii). The solution to the disc's height is performed using Brent's method. Although the disc's interior may attain vertical temperature gradients that imply convection, since our focus is on the atmospheres (which have very shallow temperature gradients, as discussed above), we ignore this complication and assume the temperature profile is exclusively set by radiation.

\subsection{Results}
\begin{figure*}
    \centering
    \includegraphics[width=\textwidth]{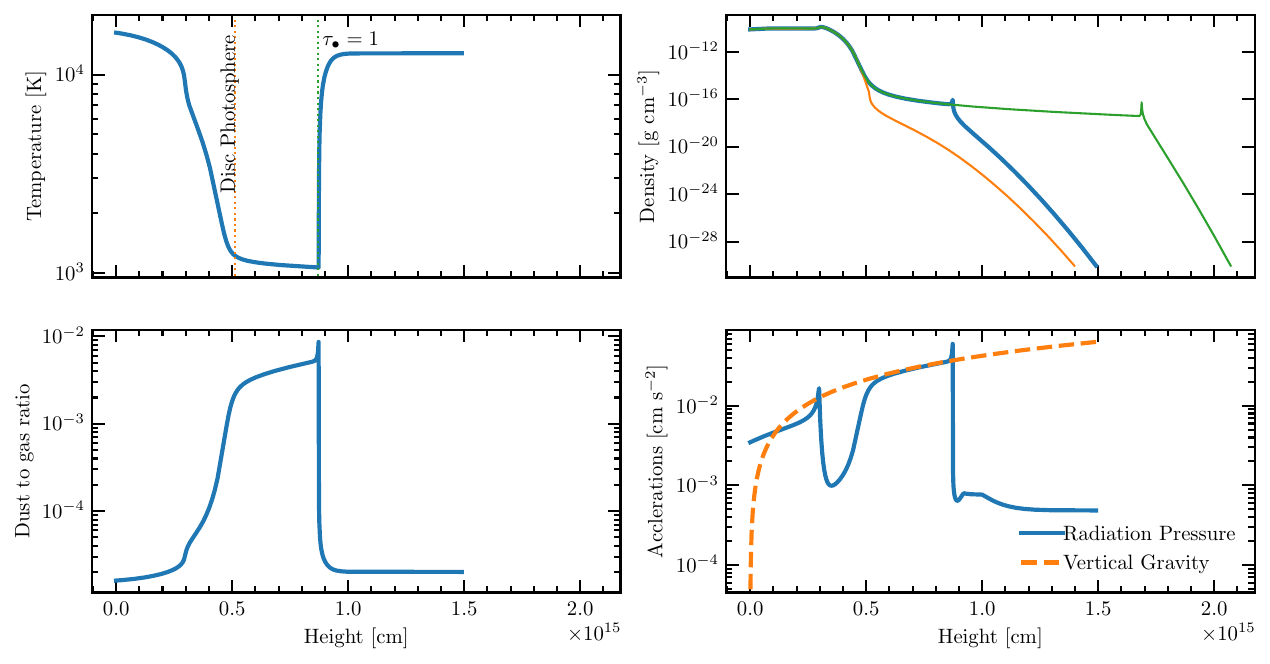}
    \caption{The vertical structure of the disc at a radius of $\sim 6.8\times10^{16}$~cm from the central black hole, with an effective temperature of $\sim 1100$~K, and flaring angle 0.003152. In the atmosphere of the disc, above the photosphere ($\tau_z \approx 0.76$), the disc is in a dynamic equilibrium between radiation pressure and vertical gravity as the dust slowly condenses with height. Once radiation from the central source is optically thin, the temperature rapidly rises, and the dust is destroyed, leading to a rapid drop in the density. In the density panel we show two additional density profiles where we increase the flaring angle to 0.02666 (orange) and lower it to 0.001516 (green). Changing the flaring angle changes the amount of radiation from the central source that is absorbed locally, which has a dramatic impact on the size of the radiation pressure supported atmosphere. }
    \label{fig:deafult_profile}
\end{figure*}

We focus our attention on regions of the disc where the effective temperature is low enough to allow for dust condensation. As our nominal model, we pick a $M_\bullet=10^8$~M$_\odot$ central black hole accreting at $\dot{M}=0.25~$M$_\odot$~yr$^{-1}$. In Figure~\ref{fig:deafult_profile} we show the vertical density and temperature profile for a radius of $6.78\times10^{16}~$cm (effective temperature 1091~K) with a central source luminosity of $\dot{M}c^2/12$, flaring angle $\mu=0.003152$, and adopting gas that has a metallicity of $Z=3Z_\odot$.

As expected for an actively heated, strongly irradiated disc, the mid-plane and the surface layers are hot. The mid-plane region varies between regions of radiation pressure and thermal pressure support. However, the temperature and density drop rapidly as one approaches the surface layers, and the radiation pressure support is minimal in this region because of the low opacity. However, in the disc's atmosphere above the photosphere to the outgoing thermal radiation, radiation pressure support due to dust begins to dominate. In the atmosphere, the disc enters the region discussed in Section~\ref{sec:smooth}, where radiation pressure and vertical gravity are essentially perfectly balanced, as the dust slowly continues to condense. In this region, the temperature and density profile become nearly flat in a region that extends across $\sim$ 20 thermal pressure scale heights. We also see a small density inversion where dust condensation finishes, and the dust is unable to continue to condense and increase the opacity, which naturally produces a density inversion to maintain dynamical balance. As discussed in Section~\ref{sec:smooth}, this is not unexpected since the opacity becomes density independent once all the dust has condensed. 

In the density panel of Figure~\ref{fig:deafult_profile}, we show the important role of the flaring angle in setting the size of the radiation pressure-supported region. Increasing the flaring angle, increases the amount of radiation from the central source that is absorbed by the disc locally. This increase in heating, increases the disc's temperature, sublimating the dust, allowing the radiation from the central source to penetrate deeper. The deeper penetration should reduce the flaring angle locally, allowing the disc to obtain a natural global equilibrium, if it were to be stable (which we actually show isn't the case in Section~\ref{sec:instability}). However, our calculations highlight the extreme sensitivity of the height of the radiation pressure-supported disc atmosphere to the flaring angle. Small changes in the flaring angle can cause this region to completely disappear (green line in Figure~\ref{fig:deafult_profile}) or almost triple in size (orange line in Figure~\ref{fig:deafult_profile}). Although exploring the global structure is beyond our 1D approach presented here, these inferences allow us to speculate about the role the radiation from the central source plays in Section~\ref{sec:central_radiation_discuss}.

\section{Stability of AGN atmospheres}\label{sec:instability}
We have seen that dust condensation in the atmospheres of AGN discs does not occur at a single height. Rather, it occurs over an extended region of many pressure scale heights where radiation pressure and vertical gravity are essentially perfectly balanced. As the height in the disc increases, the strength of gravity increases along with radiation pressure as more dust is condensed at the lower temperatures and densities that are present further up in the atmosphere. Thus, we must ask whether this balance between vertical gravity and radiation pressure is stable. 

\begin{figure}
    \centering
    \includegraphics[width=\columnwidth]{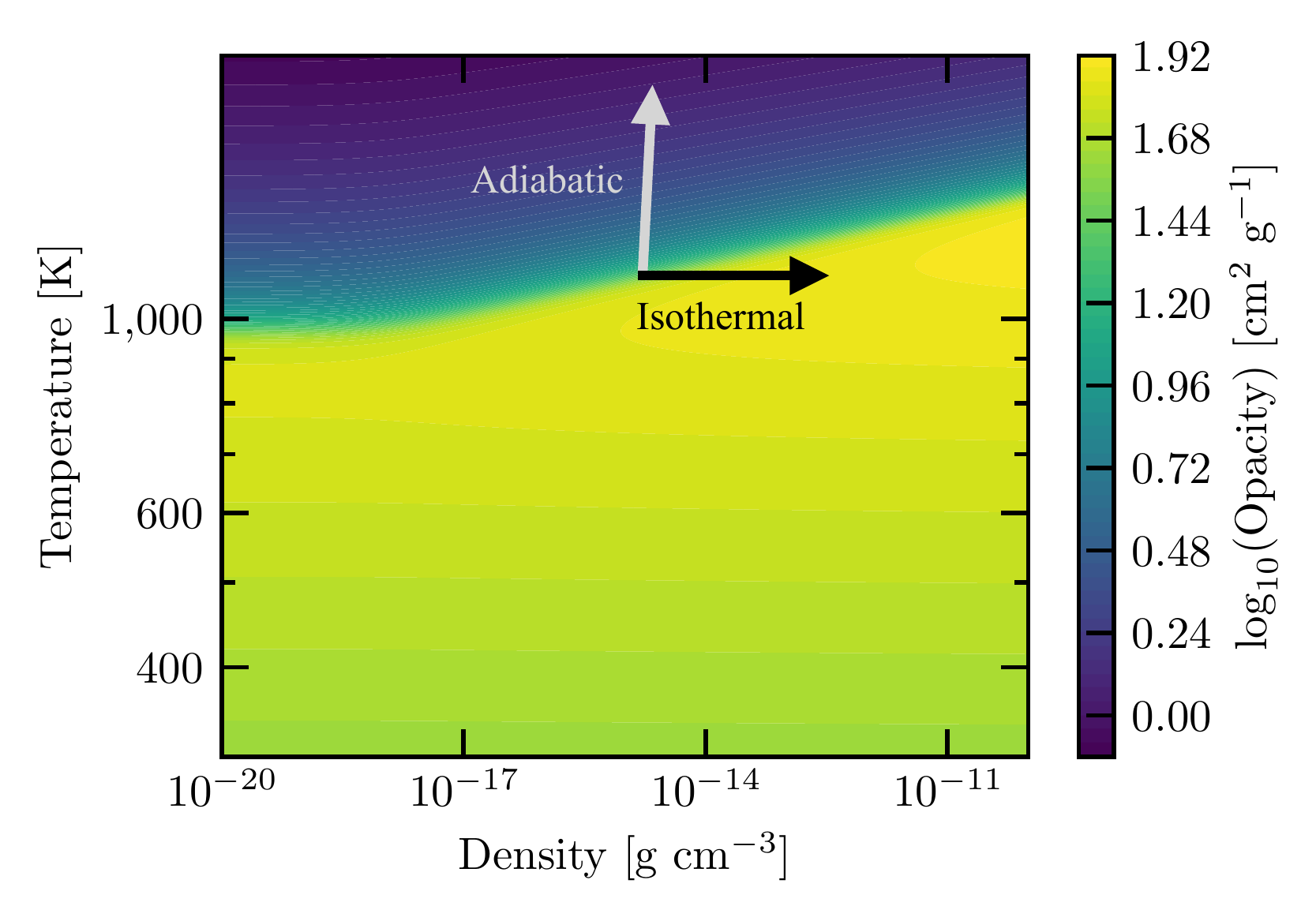}
    \caption{The opacity as a function of temperature and density with isothermal and adiabatic density perturbations shown, isothermal ones have increasing opacity, adiabatic have decreasing opacity.}
    \label{fig:simple}
\end{figure}

    We can begin to understand whether the disc's atmosphere is stable by considering a simple picture. We take a point in the disc's atmosphere where dust condensation has begun, but not completed (so that the dust fraction is density dependent), and then increase the gas density through a small perturbation. What occurs depends on the thermodynamics of the perturbation. In the case of an adiabatic perturbation (the perturbation timescale is fast compared to the local radiative timescale), then the density increase is associated with a temperature increase, e.g. $\delta T /T \propto (\delta \rho / \rho)^{(\gamma-1)}$. Alternatively, when the radiative timescale is fast, the perturbation is isothermal (in an optically thin region), and the temperature remains constant. We show in Figure~\ref{fig:simple} how these different profiles impact the change in opacity of our perturbed region. In the case of an adiabatic perturbation, the associated temperature increase results in a lower opacity as more dust is sublimated. Alternatively, in the case of an isothermal perturbation, the opacity rises. This occurs because the sublimation rate is essentially independent of the gas density, but the condensation rate increases approximately linearly with the gas density. Therefore, an increase in the gas density at a fixed temperature results in increased condensation relative to sublimation and, thus, a higher dust abundance. A higher dust abundance will result in an increase in the radiation pressure force and an upward acceleration. Since a vertically accelerated perturbation is moving into lower-density regions, it can continue unabated. 

In order to determine whether such a perturbation in the disc's atmosphere is closer to isothermal or adiabatic, we must consider the atmosphere's radiative properties.  The atmosphere is optically thin, both to the disc's emission and its own cooling radiation. The heating rate per unit mass of a patch of gas in the disc's atmosphere is thus:
\begin{equation}
    \Gamma = \kappa\left(T_{\rm eff}\right) F
\end{equation}
and the cooling rate of the gas at a temperature, $T$, is
\begin{equation}
    \Lambda = 4 \kappa\left(T\right) \sigma_{\rm SB} T^4
\end{equation}
with $\sigma_{\rm SB}$ the Stefan-Boltzmann constant. Since $T\sim T_{\rm eff}$ and $\kappa(T) \sim \kappa(T_{\rm eff})$ then the thermal timescale of this patch of gas is:
\begin{equation}
    t_{\rm th} \sim \left(\frac{k_b}{\mu_g m_h}\right) \left(\frac{1}{\kappa \sigma_{\rm SB} T_{\rm eff}^3}\right) \sim 1000\,{\rm s}
\end{equation}
for $T_{\rm eff}\sim 1000~K$ and an opacity of a few cm$^2$~g$^{-1}$. Therefore, the atmosphere of the disc will reach thermal equilibrium quickly compared to the dynamical timescale. Thus, unless the perturbation is extremely rapid, any perturbation in the AGN disc's atmosphere is likely to behave approximately isothermally. 

Therefore, this intuitive picture implies that AGN atmospheres, with an effective temperature around the dust condensation temperature, will be unstable to isothermal density perturbations.

\subsection{Basic linear (in)stability}
Our intuitive picture above implies that AGN atmospheres supported by radiation pressure on dusty gas might be unstable. To demonstrate this explicitly, we consider a simple linear stability problem before moving on to a more detailed linear analysis. 

Initially, we just consider the vertical direction and neglect thermal pressure (since it is sub-dominant in the radiation pressure-supported atmosphere). For plane-wave perturbations of the form $\delta X \sim \delta X \exp\left(-i\sigma t + ik_zz\right)$ (with $\sigma$ the frequency, $t$ time, and $k_z$ the vertical wavenumber), the perturbed continuity and the momentum equations give:
\begin{equation}
    -i \sigma \frac{\delta \rho}{\rho} + i k_z \delta u = 0
\end{equation}
\begin{equation}
    -i\sigma \delta u = \Omega^2z \frac{\delta \kappa}{\kappa}
\end{equation}
where in the last equation, we have used the fact that radiation pressure perfectly balances vertical gravity ($\Omega^2 z)$ in equilibrium. The opacity is a function of both density and temperature in the condensation region, so:
\begin{equation}
    \frac{\delta \kappa}{\kappa} = \left(\frac{\partial \log \kappa}{\partial \log \rho}\right)_T\frac{\delta  \rho}{\rho} + \left(\frac{\partial \log \kappa}{\partial \log T}\right)_\rho\frac{\delta T}{T} = a \frac{\delta  \rho}{\rho} + b \frac{\delta T}{T}
\end{equation}
Since the atmosphere is optically thin, the flux remains constant during the perturbation. Adopting local thermal balance throughout the perturbation due to the rapid thermal timescale, there is no temperature perturbation; this is because an increase in (dust) density affects both the ability of dust to absorb and emit in equal measure. 

This means that we can write the opacity change due to density as:
\begin{equation}
    \frac{\delta \kappa}{\kappa}  = a \frac{\delta \rho}{\rho}  
\end{equation}
which yields a dispersion relation:
\begin{equation}
    \sigma^2 = i \Omega^2z a k_z
\end{equation}
or, for a positive real component of $\sigma$
\begin{equation}
    \sigma = \Omega |\sqrt{a}| \sqrt{\frac{kz}{2}} + i \Omega |\sqrt{a}| \sqrt{\frac{kz}{2}} 
\end{equation}
where the imaginary part of $\sigma$ is positive for $a > 0 $ and negative for $a <0$. Thus, there is a linear instability for $a>0$. Since $\left({\partial \log \kappa}/{\partial \log \rho}\right)_T$ is order-unity and positive only in the vicinity of the condensation region (Figure~\ref{fig:adibatic}), due to the dependence increase in condensation with increasing density but fixed sublimation. Therefore, as we sketched out, an AGN atmosphere is linearly unstable to perturbations. The dispersion relation has a $\sqrt{k_z}$ dependence similar to gravity waves; however, they are longitudinal. This means that the fastest-growing modes are at the smallest scales, which are most likely to break the isothermal perturbation assumption. Thus, in the next section, we explore the instability through more realistic heating and cooling methods. Noting that the fastest-growing mode will likely be the shortest wavelength mode that can remain approximately isothermal.     

\subsection{A more detailed approach}\label{sec:detail_linear}
In the preceding discussion we identified the basic instability of radiation-pressure-supported AGN atmospheres. The key result is that this is a compressible instability that occurs for fixed temperature. Since rapid compressible perturbations are adiabatic, we must consider the effects of compressibility and thermal relaxation, as well as whether it is manifestly 1D (at least on small scales). For simplicity, we still adopt the plane-wave approach, which we extend to 3D perturbations. In the vertical direction, this only requires $k_z z > 1$ as the density and temperature profiles are slowly varying in the atmosphere with height. However, in the radial and azimuthal directions it is stricter, neglecting curvature and shear requires that $k_RH>1$ and $Rk_\phi H > 1$. As we will show, the fastest growing modes have $kH\gg 1$. Thus, we can neglect the complications of shear and curvature here and proceed with a Cartesian approach, where we neglect background gradients in both the radial ($\mathbf{\hat{x}}$) and vertical directions. 

\begin{figure}
    \centering
    \includegraphics[width=\columnwidth]{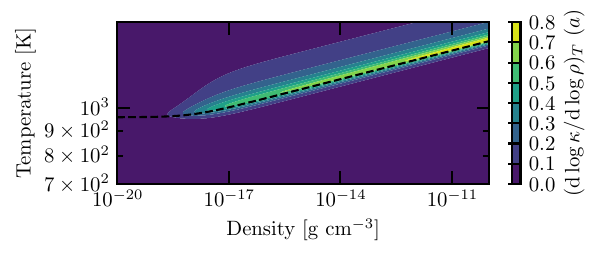}
    \includegraphics[width=\columnwidth]{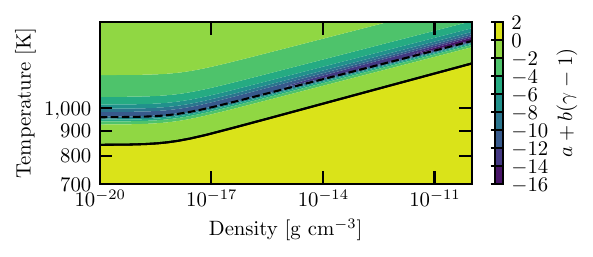}
    \caption{The value of $a$ (top), and $a+b(\gamma-1)$ (bottom, for $\gamma=7/5$), the adiabatic term in the small $k$ dispersion relationship. The black solid line is the point where $a+b(\gamma-1)$ equals zero, the dashed line is the position of dust condensation. We see that the dust condensation region is unstable to isothermal perturbations ($a>0$), but stable to adiabatic perturbations ($a+b(\gamma-1)<0)$.}
    \label{fig:adibatic}
\end{figure}

We reintroduce the thermal pressure into our dynamic equations and consider the evolution of energy (pressure) through a Newtonian cooling approach. Thus, the 3D fluid equations are:
\begin{eqnarray}
\frac{\partial \rho}{\partial t} + \nabla\cdot(\rho \mathbf{u}) &=&0\\
\frac{\partial \mathbf{u}}{\partial t} +\mathbf{u}\cdot \nabla \mathbf{u} &=& -\frac{1}{\rho}\nabla P + \frac{1}{c}\kappa F\mathbf{\hat{z}} +\mathbf{g} \\
\frac{\partial P}{\partial t} + \mathbf{u}\cdot\nabla P &=& -\gamma P \nabla \cdot \mathbf{u} - \frac{P}{T}\frac{T-T_{\rm eq}}{\tau_{\rm th}}
\end{eqnarray}
where $\mathbf{u}$ is the velocity, $\mathbf{g}$ is the gravitational acceleration, $\gamma$ is the ratio of specific heats, $T_{\rm eq}$ is the temperature in thermal equilibrium, and $\tau_{\rm th}$ is the (assumed) constant thermal relaxation timescale. Thus, using plane-wave perturbations in our Cartesian approach, ignoring stratification, and adopting $\mathbf{g}\approx\Omega^2z$, the perturbed equations become:
\begin{eqnarray}
    -i \sigma \frac{\delta \rho}{\rho} + i k_x \delta u_x + i k_y \delta u_y + i k_z \delta u_z &=& 0\\
    -i \sigma \delta u_x + i k_x \frac{\delta P}{\rho} &=& 0 \\
    -i \sigma \delta u_y + i k_y \frac{\delta P}{\rho} &=& 0 \\
    -i \sigma \delta u_z + i k_z \frac{\delta P}{\rho} -\frac{\delta \kappa}{\kappa} \beta \Omega^2 z &=&0\\
    -i \sigma \frac{\delta P}{\rho} + i\gamma c_s^2\left(k_x \delta u_x + k_y \delta u_y + k_z \delta u_z \right)\nonumber\\ + \frac{1}{\tau_{\rm th}}\frac{\delta P}{\rho} - \frac{c_s^2}{\tau_{\rm th}}\frac{\delta \rho}{\rho} & =&  0
\end{eqnarray}
where $k_x$ \& $k_y$ are the wavenumbers in the $\mathbf{\hat{x}}$ and $\mathbf{\hat{y}}$ (azimuthal) directions, respectively, $c_s=\sqrt{k_b T / \mu_g m_h}$ is the isothermal sound speed of the background state and $\beta$ is the ratio of the strength of the radiation pressure to vertical gravity. Now writing the opacity perturbation as:
\begin{equation}
    \frac{\delta \kappa}{\kappa} = \left(\frac{\partial \log \kappa}{\partial \log \rho}\right)_T\frac{\delta  \rho}{\rho} + \left(\frac{\partial \log \kappa}{\partial \log T}\right)_\rho\frac{\delta T}{T} = \left(a-b\right)\frac{\delta \rho}{\rho} + \frac{b}{c_s^2}\frac{\delta P}{\rho}\,,
\end{equation}
 the perturbed equation for the vertical momentum becomes:
\begin{equation}
    -i \sigma \delta u_z + i k_z \frac{\delta P}{\rho} -(a-b) \beta \Omega^2 z\frac{\delta \rho}{\rho} -\frac{b}{c_s^2}\beta \Omega^2 z\frac{\delta P}{\rho} =0\,.
\end{equation}
The solution of these equations yields the following dispersion relation:
\begin{eqnarray}
    \sigma^2\Bigg\{i \sigma^3 &-& \sigma^2\frac{1}{\tau_{\rm th}} - i k^2 \sigma \gamma c_s^2 + k_z\sigma \left[a + b\left(\gamma -1\right)\right]\beta \Omega^2 z\nonumber \\ &+& i k_z \left(\frac{a\beta \Omega^2 z}{\tau_{\rm th}}\right) + k^2\frac{c_s^2}{\tau_{\rm th}}  \Bigg\}= 0 
\end{eqnarray}
where $k^2=k_x^2+k_y^2+k_z^2$, where we can ignore the trivial $\sigma = 0$ solution. First, we consider the adiabatic limit $\tau_{\rm th}\rightarrow \infty$, which yields the dispersion relation:
\begin{equation}
    i\sigma^2 - i k^2\gamma c_s^2 + k_z\left[a + b\left(\gamma -1\right)\right]\beta \Omega^2 z = 0
\end{equation}
This clearly returns standard adiabatic sound waves in the short-wavelength (large $k_z$) limit and an opacity term:
\begin{equation}
    \sigma^2 = i k \left[a + b\left(\gamma -1\right)\right]\beta \Omega^2 z
\end{equation}
where the imaginary component of $\sigma$ depends on the sign of $a + b\left(\gamma -1\right)$. Since $b$ is large and negative, this term is always negative in the condensation region (see Figure~\ref{fig:adibatic}). Thus, the atmosphere is stable to adiabatic perturbations. 
We now consider the opposite limit $\tau_{\rm} \rightarrow 0$, namely the limit of rapid cooling (essentially the isothermal case). In this case, the dispersion relation becomes:
\begin{equation}
    \sigma^2 -ik_z a\beta\Omega^2z - k^2c_s^2 =0
\end{equation}
Again in the short-wavelength limit, we have (isothermal) sound waves, but in the long-wavelength limit (small $k$), we return to exactly the same dispersion relation discussed in the simple problem above, which we have demonstrated can become positive and order unity during dust condensation (Figure~\ref{fig:opac_index}). 

\begin{figure*}
    \centering
    \includegraphics[width=0.495\textwidth]{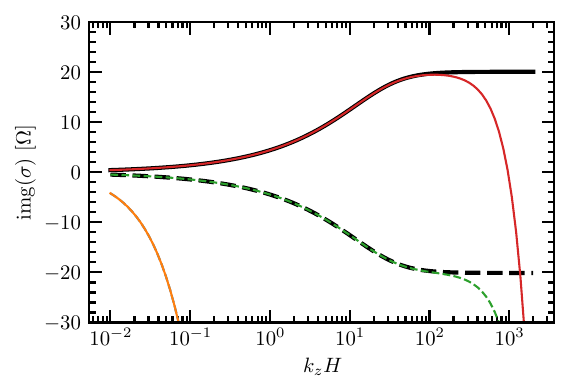}
    \includegraphics[width=0.495\textwidth]{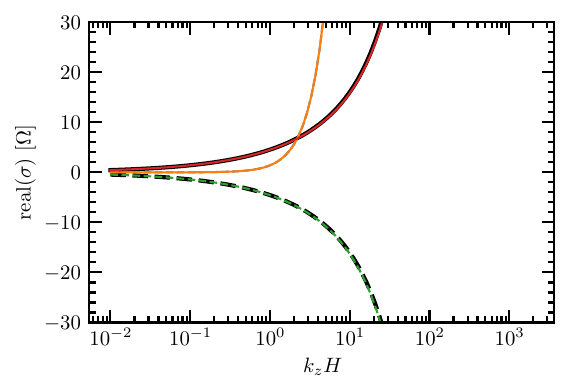}
    \caption{An example numerical solution to the dispersion relation shown in red solid (unstable) and green dashed (stable), the black solutions show the pure isothermal solutions (solid: unstable, dashed: stable), and the orange shows the pure adiabatic solution. The full solution is well approximated by the isothermal approximation to very short wavelengths.}
    \label{fig:numerical}
\end{figure*}

We can now consider the full solution to the dispersion relation. Turning the dispersion relation into dimensionless form by normalising the frequency by $\Omega$ and $k$/$k_z$ by the effective disc scale height ($H=c_s/\Omega$), we find:
\begin{equation}
    i\sigma^3 -\frac{\sigma^2}{\tau_{\rm th}}-ik^2\sigma\gamma + k_z\sigma\left[a + b\left(\gamma -1\right)\right]\beta \left(\frac{z}{H}\right) +\frac{ik_za\beta}{\tau_{\rm th}}\left(\frac{z}{H}\right) +\frac{k^2}{\tau_{\rm th}}=0
\end{equation}
where $\sigma \rightarrow \sigma/\Omega$,  $k\rightarrow kH$, $k_z\rightarrow k_zH$, and $\tau_{\rm th}\rightarrow\tau_{\rm th}\Omega$. We plot the solution to the dispersion relation for typical parameters (with $k=k_z$) for $\tau_{\rm th}=10^{-4}$, $z/H=58$, $a=0.7$, $b=-20$ and $\beta=0.99$ in Figure~\ref{fig:numerical}. This demonstrates that for rapid cooling timescales, as expected in the disc’s atmosphere, an approximately isothermal, linearly unstable mode is a good approximation.. 
\section{Hydrodynamic Simulations}\label{sec:simulations}
To understand the non-linear evolution of this instability, we appeal to hydrodynamic simulations. At this stage, to isolate the physics of our new instability, we focus on high-resolution two-dimensional ($R-z$) simulations of the disc's atmosphere. Focusing on this region allows us to significantly simplify the radiative transfer in the simulation since the material above the disc's photosphere is generally optically thin compared to the outgoing IR radiation. Furthermore, like in our linear analysis, we consider small radial scales and ignore the rotation (and shear) in the background disc. Therefore, we choose to simulate a narrow 2D Cartesian box placed in the disc's atmosphere.

\subsection{Methods}
It is tempting to run the simulations with a modern shock-capturing code (e.g. Athena++, \citealt{athena_code} or PLUTO, \citealt{pluto_code}). However, given our instability originates in an initially static atmosphere where gravity balances radiation and thermal pressure exactly, such an approach would be a poor choice. Shock capture codes are well known to be unable to initially preserve static profiles under gravity \citep{Kappeli2014}. This would mean that we would be unable to be certain that any instability we identify would arise from our physical instability rather than from numerical effects without a great deal of effort and excess resolution. Although these numerical issues can be fixed with ``well-balancing'' corrections \citep[e.g.][]{Kappeli2016}, we choose a different and simpler approach. Instead, we use the {\sc zeusmp}V2 code \citep{Hayes2006}. Since, in the {\sc zeus} approach, forces and transport are operator split, and it makes use of a staggered mesh \citep{Stone1992}, it is possible to construct an initial setup that is in perfect numerical equilibrium. Thus, any instability that develops from this initial state can be identified as a physical instability. Thus, while we acknowledge that we are sacrificing excess diffusivity (compared to a shock-capturing code), we can be confident that we are correctly identifying the physical instability. Indeed, in future work, since we successfully simulate the instability here,  one could make use of a shock-capturing code. 

\subsubsection{``Radiation Transport'' and Thermodynamics}
Given we have deliberately chosen to simulate only the upper atmosphere where the outgoing disc's IR radiation is optically normally thin, we can consider the disc's emission to have a mean intensity that is constant in time. Essentially we are assuming that it is only the optically thin region (to the disc's emission) of the atmosphere that is dynamic, and that the photosphere to the outgoing radiation and the disc region below the photosphere remain static. Furthermore, we are also assuming that the external irradiation does not appreciably change the mean intensity below its photosphere (a result true in our static calculations). 

Therefore, to proceed with our thermal calculation, we take the mean intensity of the thermal emission of the disc ($J$) derived from our 1D numerical calculations of the vertical structure of the disc and interpolate this onto our hydrodynamic grid and assume that it remains constant in time. This approach means that we do not need to perform on-the-fly radiative transfer for the disc's thermal emission. To incorporate radiation from the central source, similar to our 1D numerical calculations, we perform a ray-tracing calculation along the z coordinate to determine $\tau_z$; we do this every time step for every radial grid spacing. We then fix $\mu$ and $\gamma'$ in the simulation to determine $\tau_{\rm irr}$ and calculate the irradiation heating from Equation~\ref{eqn:irradiation_heating}. As we describe in the discussion, fixing $\mu$ implies that the flaring angle does not evolve significantly or permit shadowing, effects that we will clearly see develop in our simulation. However, this approach does allow the optical depth to radiation from the central source to respond dynamically to the density field. 

Since the thermal timescale is short compared to the dynamical timescale of the simulation in the disc's atmosphere, we adopt a locally isothermal approach where the temperature is solved for at every time step through Equation~\ref{eqn:energy_balance} and the pressure is determined through the ideal gas law by updating the isothermal sound-speed locally every time step, such that:
\begin{equation}
    P(R,z,t)=\rho(R,z,t)c_s^2(R,z,t)
\end{equation}

In our simulations, we use the same opacity and dust model as used in our 1D static models, where the opacity is updated before the source step in the {\sc zeus} algorithm.  

\subsubsection{Construction of an initially numerically static profile}
One of the advantages of the {\sc zeus} algorithm is that it allows the construction of initial conditions that are in perfect numerical equilibrium with gravity, provided that gravity aligns with one of the coordinate axes. We implement gravity in the vertical direction $g_z = \Omega^2 z$ to match our 1D solutions, noting that the $z/R<1$ criteria is not violated in our simulation boxes.

To set up our initial static profile, we proceed as follows: given our flux $F_z$, we implement radiation pressure on the cell boundaries (a-grid in {\sc zeus} terminology). To do this, we insert an extra acceleration during the source step in the $\mathbf{\hat{z}}$ direction of the form:
\begin{equation}
    a^z_{{\rm rad},i,j} = \frac{1}{2}\left[\kappa(\rho_{i,j},T_{i,j})+\kappa(\rho_{i,j-1},T_{i,j-1})\right]\frac{F_z}{c}
\end{equation}
where $i$ and $j$ are the cell indices in the radial and vertical directions, respectively. Given that the mean intensity, $J$, is spatially fixed for all time, we then interpolate our density profile from the 1D simulations to find the density in the first ghost cell in the inner boundary. We then numerically solve the {\sc zeus} source force step from the inner boundary to set up an initial density profile that is in perfect dynamical balance between gravity, thermal pressure, and radiation pressure. This is verified by advancing the code one time step forward and noting the source step generates zero velocity, producing no change in the transport step. Thus, to see the instability, we perturb our initial condition with $(1 + A \sin^2)$ perturbation in both $R$ and $z$ with an amplitude of $A=10^{-5}$ and wavenumber of $10^{13}$~cm. We find our results in both the linear phase and non-linear phase are independent of this choice. In the inner vertical boundary, we perform the same procedure inwards to set the values in the ghost cells, which we then fix for the remainder of the simulation. Along the radial boundaries, we apply periodic boundary conditions and use a diode outflow boundary condition (which permits outflow but not inflow) at the upper edge of our box.  Finally, we apply a density floor of $10^{-35}$~g~cm$^{-3}$ to the simulation box, without this, the upper region of the box at extremely low density can result in underflow errors. This density floor is orders of magnitude lower than the density in regions of interest, and our results are insensitive to this choice. 

\subsubsection{Simulations}
At this initial stage, to build up an understanding of how this instability operates and ultimately results in a ``failed wind'' we run three simulations at different radial distances, one just at the point where dust begins to condense in the disc's atmosphere ($R=4\times10^{16}$~cm), resulting in a puffed up radiation pressure-supported atmosphere and a further two at $2.5\times$ and $5\times$ further out in radius. In the simulations, we set the base of the simulation in the vertical direction to be just below the photosphere. This construction means that at the lower boundary of our simulation we have a disc region that is essentially dust-free and supported by thermal pressure. We run our simulations with a CFL number of $0.5$, and a standard artificial viscosity parameter \texttt{qcon} of $2.0$. 

Our standard resolution is $N_R\times N_z$ of 576$\times$2048 and we enforce square cells. The grid placement is detailed in Table~\ref{tab:sim_table}. The vertical extent of the simulation is chosen so that the ``failed wind'' remains fully within the simulation domain. Although our simulations are fairly high resolution (at 2048 cells in the vertical direction), due to the fact that the pressure scale height in the dusty atmosphere is extremely small compared to the box heights required, the resolution to the pressure scale height in the dusty region is $\sim 5-10$. This does not mean that we are under-resolving our initial state as the pressure scale height in the underlying hot disc is significantly larger, and the dusty region is radiation pressure supported with a much larger pressure scale height. However, this means that since our linear instability is predicted to grow with a rate that is $\propto \sqrt{k}$, our resolution will set the fastest growing mode. Furthermore, physically resolving the expected fastest growing mode, $k_zH\sim 100$, from our linear analysis (Figure~\ref{fig:numerical}) is infeasible. Thus, as we detail in the discussion, this has some implications for how we interpret our simulations in comparison to the observations.

The simulations are then run for several tens of dynamical timescales after the nonlinear state becomes pseudo-steady, where the time-averaged RMS velocity in the disc's atmosphere has converged. It typically takes $\sim 20$ dynamical times for this pseudo-steady state to be reached.  
\begin{table}
    \centering
    \begin{tabular}{c|c|c|c}
        & $R$ & $Z_{\rm min}$ & $Z_{\rm max}$ \\
        \hline
        Inner R & $4\times10^{16}$~cm & $4.75\times10^{14}$~cm & $1\times10^{16}$~cm \\
        Mid R & $1\times10^{17}$~cm & $7.3\times10^{14}$~cm & $4\times10^{16}$~cm \\
        Large R & $2\times 10^{17}$~cm & $2\times10^{15}$~cm & $1.5\times10^{17}$~cm \\
        \hline
    \end{tabular}
    \caption{Dimensions of the simulation boxes}
    \label{tab:sim_table}
\end{table}

\subsection{Results}
\begin{figure*}
    \centering
\includegraphics[width=0.495\textwidth]{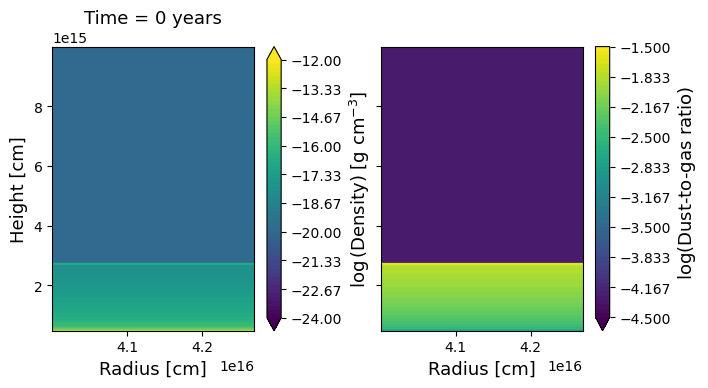}
    \includegraphics[width=0.495\textwidth]{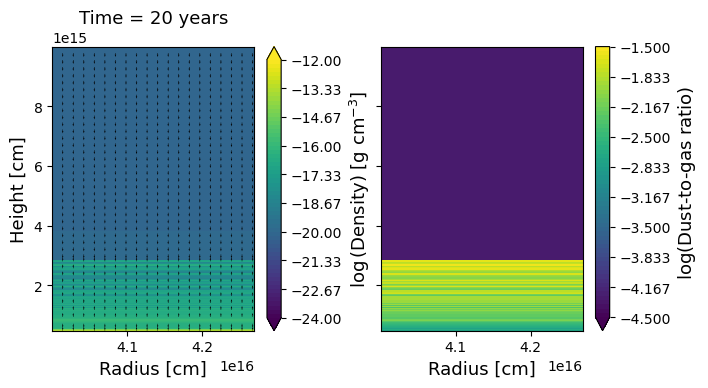}
    \includegraphics[width=0.495\textwidth]{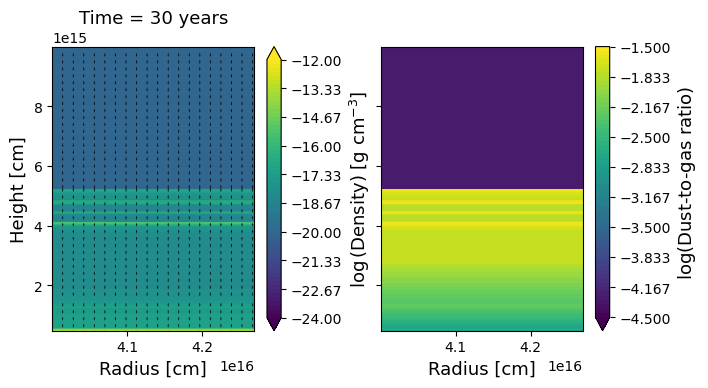}
    \includegraphics[width=0.495\textwidth]{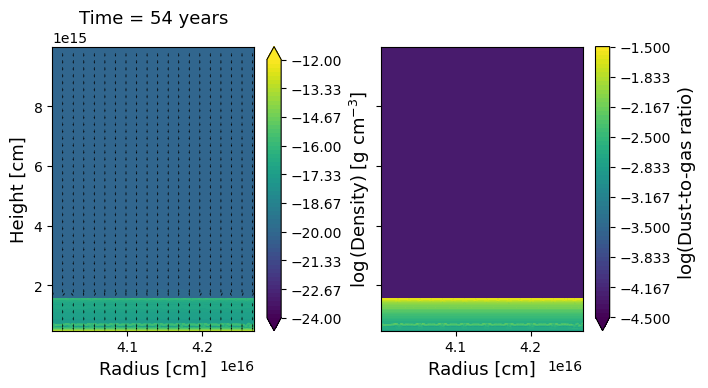}
    \includegraphics[width=0.495\textwidth]{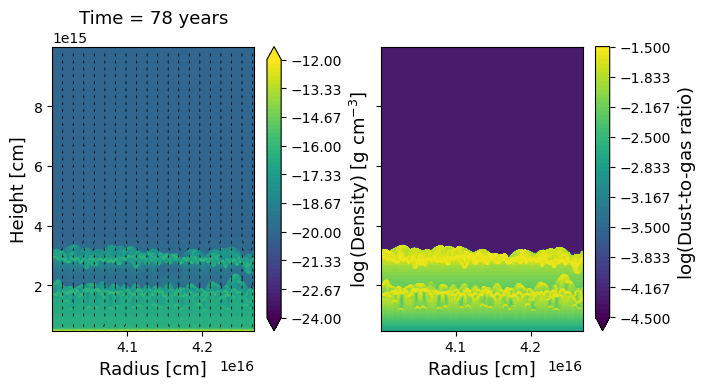}
    \includegraphics[width=0.495\textwidth]{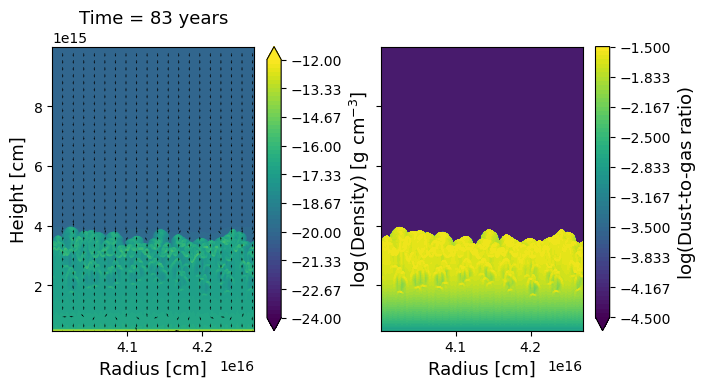}
    \includegraphics[width=0.495\textwidth]{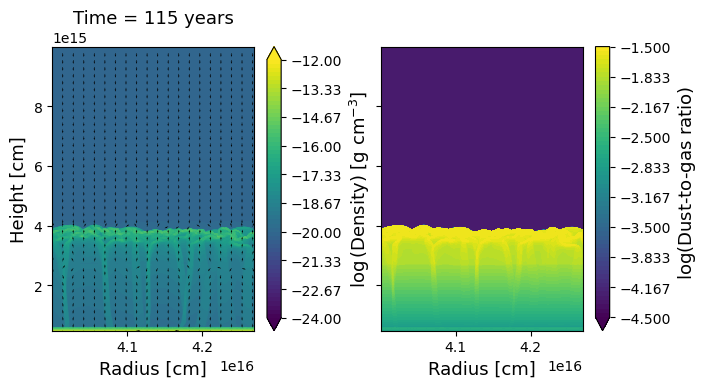}
    \includegraphics[width=0.495\textwidth]{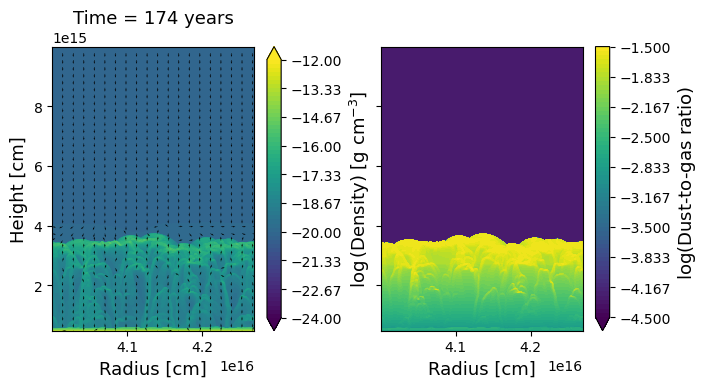}
    \includegraphics[width=0.495\textwidth]{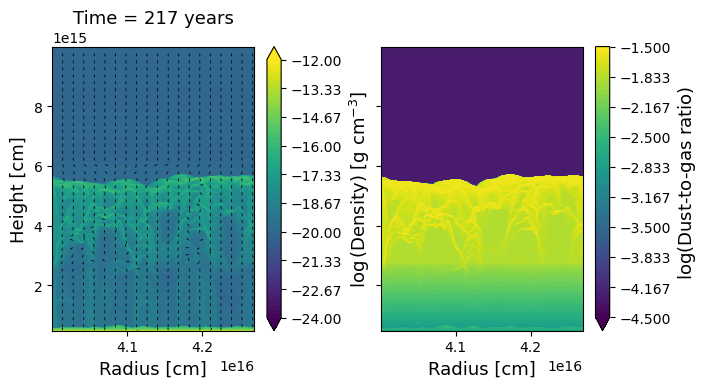}
    \includegraphics[width=0.495\textwidth]{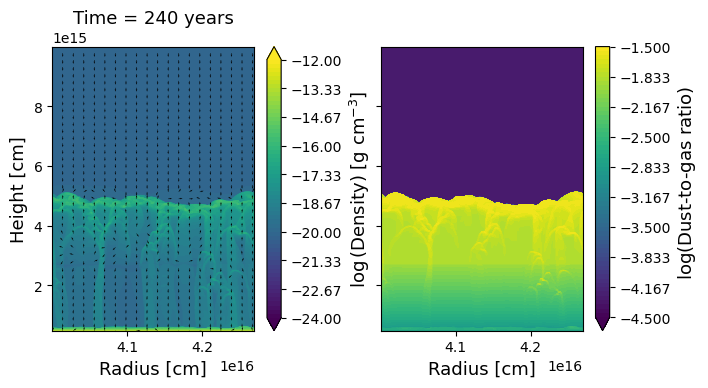}
    \caption{Snapshots of the density, velocity field (shown as normalised vectors on density field), and dust-to-gas ratio in our Inner R simulations as a function of time. A video of the simulation is at \url{https://doi.org/10.5281/zenodo.17526011}. }
    \label{fig:inner_R_sim_panels}
\end{figure*}

\begin{figure*}
    \centering
    \includegraphics[width=0.495\textwidth]{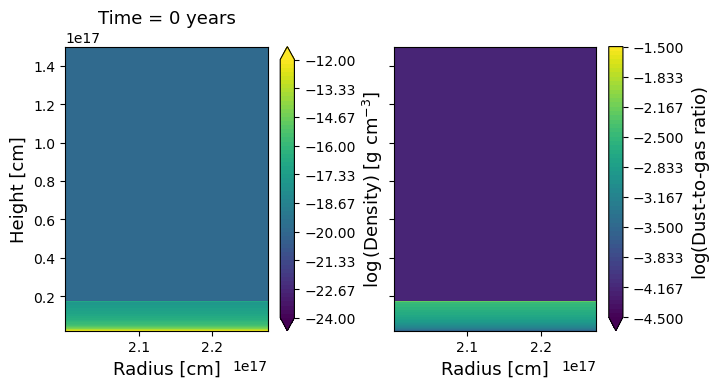}
    \includegraphics[width=0.495\textwidth]{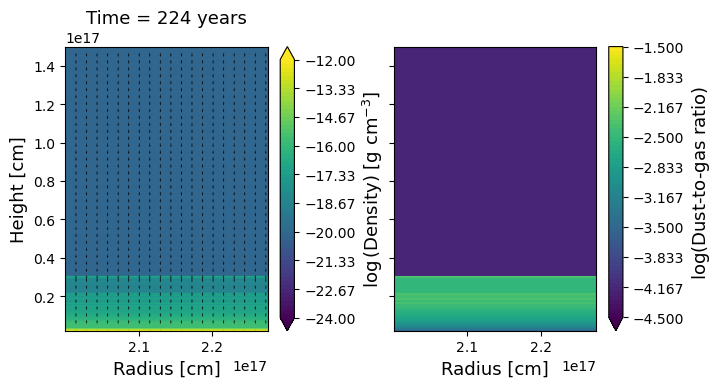}
    \includegraphics[width=0.495\textwidth]{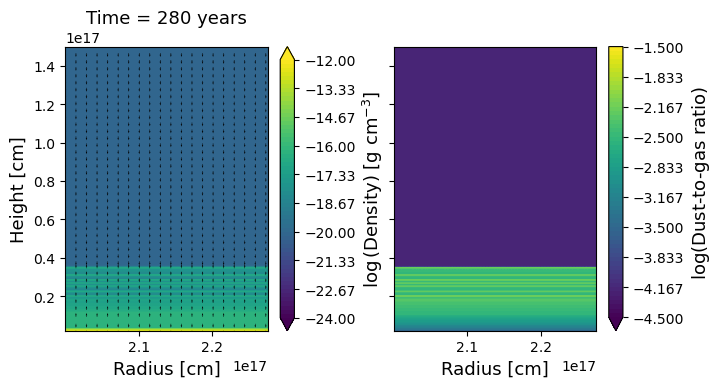}
    \includegraphics[width=0.495\textwidth]{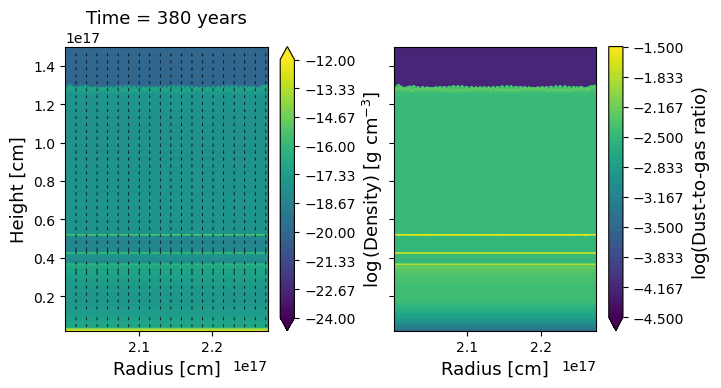}
    \includegraphics[width=0.495\textwidth]{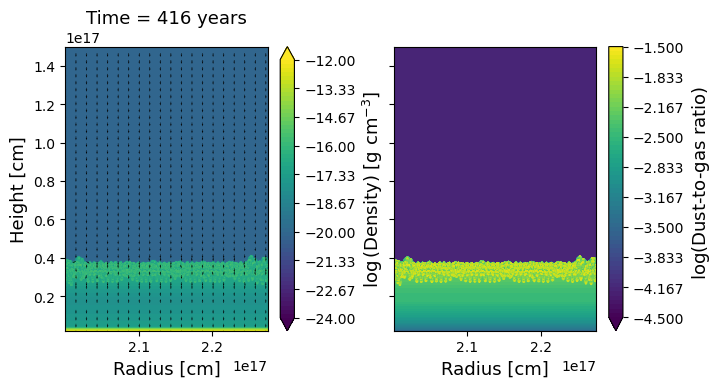}
    \includegraphics[width=0.495\textwidth]{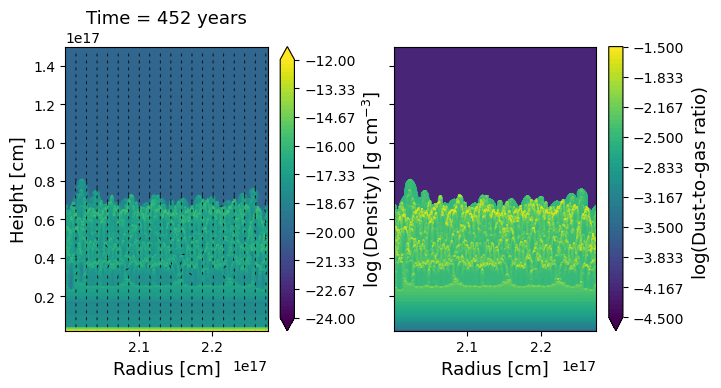}
    \includegraphics[width=0.495\textwidth]{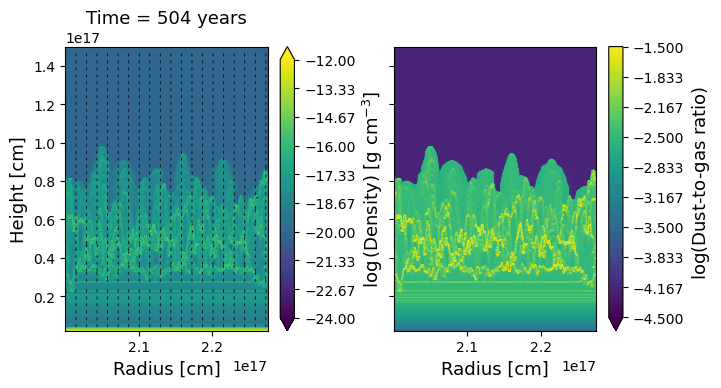}
    \includegraphics[width=0.495\textwidth]{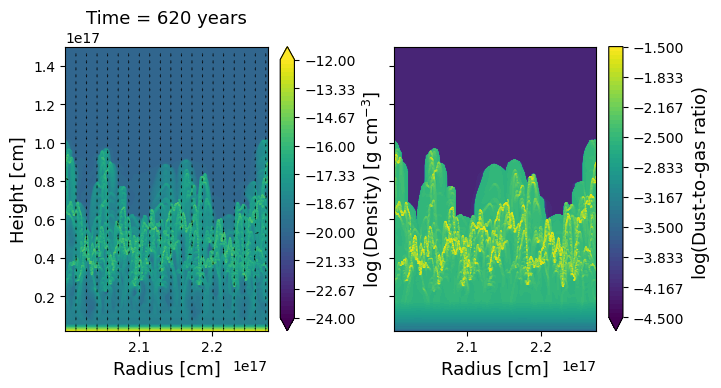}
    \includegraphics[width=0.495\textwidth]{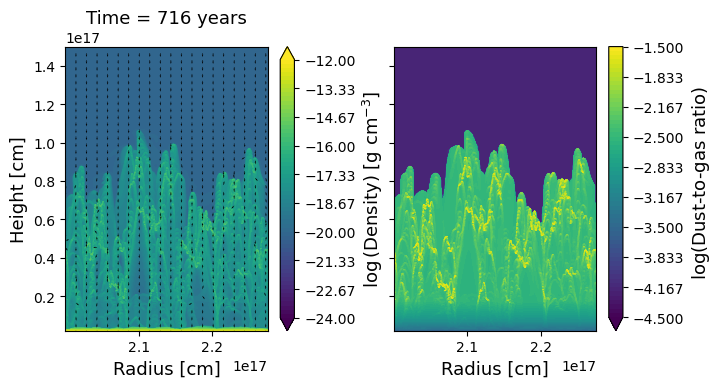}
    \includegraphics[width=0.495\textwidth]{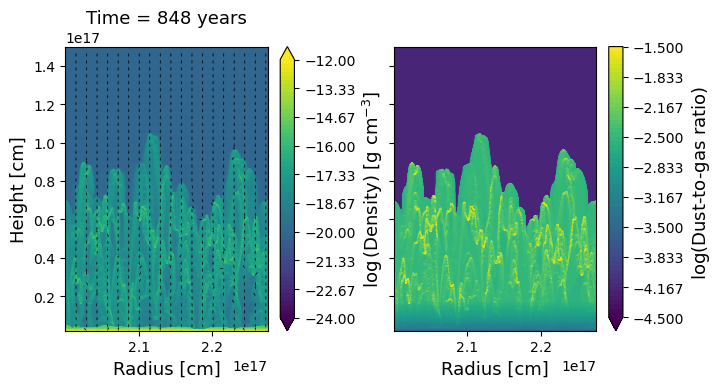}
    \caption{Snapshots of the density, velocity field (shown as normalised vectors on density field), and dust-to-gas ratio in our Large R simulations as a function of time.  A video of the simulation is at \url{https://doi.org/10.5281/zenodo.17526011}.}
    \label{fig:outer_R_sim_panels}
\end{figure*}

In Figures~\ref{fig:inner_R_sim_panels} \&~\ref{fig:outer_R_sim_panels}, we show snapshots of our simulations from the linear to nonlinear stage for both the Inner R and Large R cases. The evolution of all our simulations follows the same general stages. First, the linear instability develops in the disc's upper atmosphere; this is accompanied by a bulk movement of the entire atmosphere up and down as it is accelerated to higher levels by the radiation pressure before collapsing onto the disc again. The atmosphere then transitions to a non-linear phase where ``fountains'' of material are driven up by radiation pressure before being exposed to the radiation from the central source and the dust is sublimated. Once the dust is sublimated, the acceleration due to radiation pressure ceases; the clumps then expand as they are heated to $\sim 10^4$~K (see the discussion, where in reality radiation pressure from the central source, which we don't include, will compress the clumps). The clumps then decelerate under vertical gravity as they flow out of the disc before they eventually fall back towards the disc. We find speeds of a few hundred to $\sim$1000 km~s$^{-1}$ are reached in these fountains, as the disc atmosphere material is accelerated and launched vertically upward, and they reach a similar velocity as they crash back towards the disc. Hence, we identify these fountains of material as the clumpy ``failed'' wind.  

\begin{figure}
    \centering
    \includegraphics[width=\columnwidth]{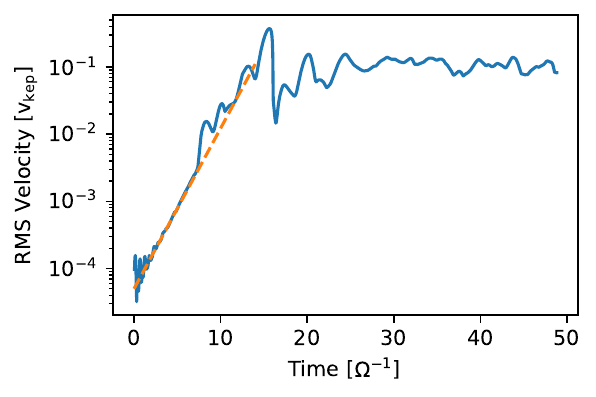}
    \caption{The evolution of the RMS velocity in the dusty region of the disc's atmosphere ($X_d>3\times10^{-3}$) for the Large R simulation. The velocity shows a linear growth phase (as indicated by the dashed line), growing on a timescale of $\sim \Omega^{-1}$. Eventually, after several dynamical timescales the RMS velocity saturates at around 10\% of the Keplerian velocity. }
    \label{fig:vrms}
\end{figure}

\begin{figure}
    \centering
    \includegraphics[width=\columnwidth]{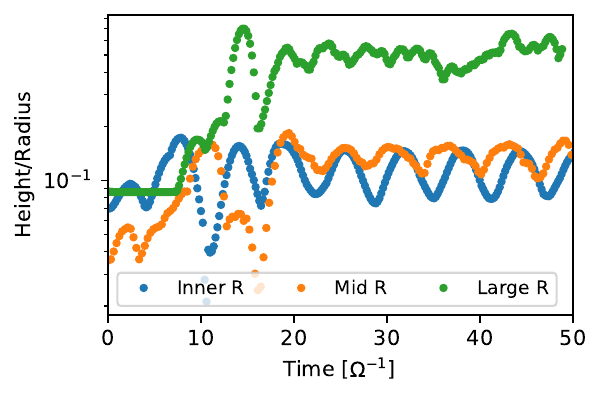}
    \caption{The time evolution of the disc atmosphere's height in our simulations for our three cases. The average height is larger than the static case and it oscillates on a timescale of $\sim 5-6~\Omega^{-1}$. }
    \label{fig:height_time}
\end{figure}

To analyse our instability in Figure~\ref{fig:vrms}, we show the evolution of the root-mean-squared velocity as a function of time in the Large R simulation. Here we restrict ourselves to regions that contain some dust ($X_d>3\times10^{-3}$). We see a linear growth phase characteristic of a linear instability as the velocity fluctuations grow from our initially seeded density profile.  The velocity fluctuations continue to grow over $\sim 20$ dynamical timescales, where after about $\sim 7$ dynamical timescales the linear growth in the velocity fluctuations is punctuated by larger amplitude oscillations arising from the bulk oscillation of the disc's atmosphere (see~Figure~\ref{fig:height_time}), before the velocity fluctuations saturate and oscillate about an approximately constant level around $\sim 10\%$ of the Keplerian velocity. The growth rate during the linear phase is $\sim 0.55 \Omega$, consistent with our expectation from our linear analysis.  

\begin{figure}
    \centering
    \includegraphics[width=\columnwidth]{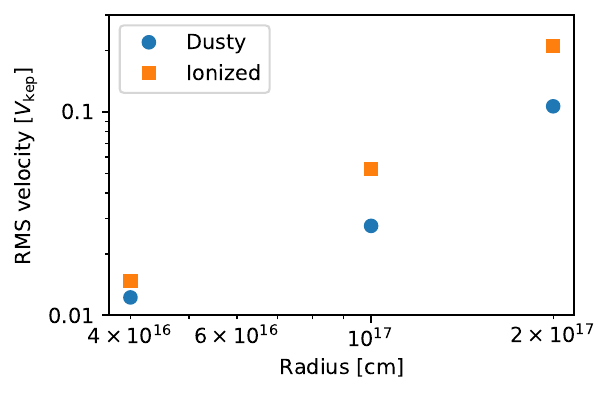}
    \caption{The RMS velocity in the dusty region of the AGN atmosphere and the ionized region as a function of distance from the central black hole.}
    \label{fig:rms_velocity_distance}
\end{figure}

We find that this saturated RMS velocity is a function of the radial distance. Figure~\ref{fig:rms_velocity_distance} shows how the RMS velocity scales as a function of distance from the super-massive black hole, both in the dusty ($X_d>3\times10^{-3}$) and ionized regions ($X_d<3\times10^{-3}$)\footnote{We also excluded the underlying disc material, and require the region has a density $10^4$ larger than the floor density.}, finding that it increases with distance. The velocity is higher in the ionised region as it is only the highest velocity regions of the dusty fountains that make it into the ionized region. Furthermore, near where the dust first condenses in the atmosphere, the RMS velocity is only a few percent of the Keplerian; however, it quickly grows to larger than 20\% in the ionized regions five times further away. 

\begin{figure}
    \centering
    \includegraphics[width=\columnwidth]{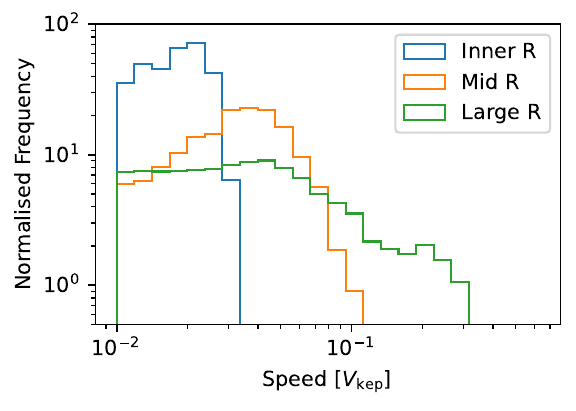}
    \caption{The distributions of speeds in our ionized region of the AGN atmosphere are shown for the three different simulation radii.}
    \label{fig:v_dist}
\end{figure}

Furthermore, in Figure~\ref{fig:v_dist}, we show the velocity distribution of the ionized clumps in our simulations for the three different radii. This shows that the distribution is not narrowly peaked but has a reasonable width (of order to the RMS velocity). However, the distribution is skewed, with a longer tail to lower speeds and a steeper fall-off at higher velocities. The dust-to-gas ratio in fast moving fountains (speeds $>$ 300 km~s$^{-1}$) is correlated with density, with the highest density clumps having the highest dust-to-gas ratios, approaching $0.01$ in densities $\gtrsim 10^{8}$~cm$^{-3}$.  We leave our discussion of the density distribution of the clumps to the discussion. 

\begin{figure*}
    \centering
    \includegraphics[width=0.33\textwidth]{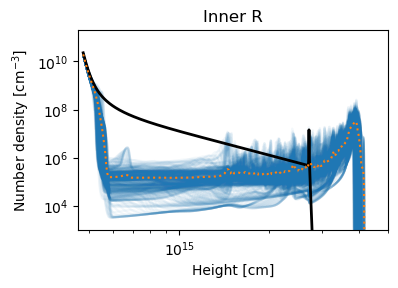}
    \includegraphics[width=0.33\textwidth]{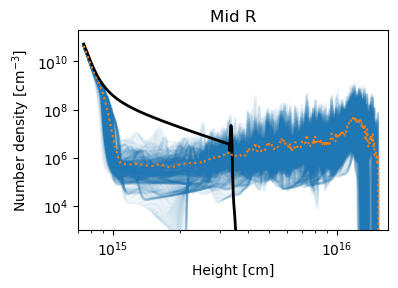}
    \includegraphics[width=0.33\textwidth]{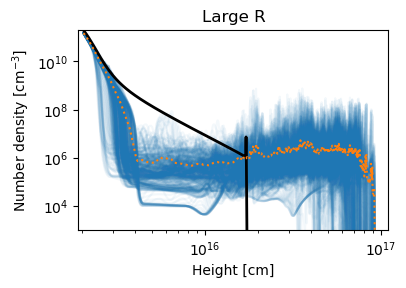}
    \caption{The initial density profile (solid, black) compared to the variation of density with height after 30 dynamical timescales (thin, blue lines) and their mean (dashed, orange), shown for our three simulation radii. The dynamic density profile is typically lower and flatter than the initial density profile. The densities spike to values around $\gtrsim 10^8~$cm$^{-3}$ in that disc atmosphere.}\label{fig:compare_densities}
\end{figure*}

Finally, we consider the impact of the instability on the density profile, compared to the static solution. In Figure~\ref{fig:compare_densities}, we show the initial density profiles of our three simulations compared to the individual density profiles after 30 dynamical timescales. 

These density profiles show that at small heights the initial density profiles track the initial conditions, indicating that the disc region without dust remains in hydrostatic equilibrium. This agreement between the static and dynamic density profile close to the inner boundary validates our assumption that ignoring the on-the-fly treatment of radiative transfer is a reasonable first approximation, and justifies our choice only simulating the disc atmosphere. The dynamic density profiles then drop rapidly below the initial conditions to around a typical value of $10^6$~cm$^{-3}$. The density profile then remains fairly flat, although slowly increasing, extending beyond the initial density profile. Finally, the density profile rapidly drops once the irradiation from the central source heats the gas, sublimating the dust. This increase in density beyond the initial profile increases with radius; however, it is variable over time. 

In Figure~\ref{fig:height_time} we track the average ``height'' of the atmosphere in each of the simulation boxes by measuring the height at which the average number density drops below 10$^2$~cm$^{-3}$ as a function of time (although our results are insensitive to this choice). This figure shows that the average height is a factor of a few higher than that calculated from the static calculations, but also that it varies up and down significantly.  After about 20 dynamical times the height variations have settled into an oscillation pattern which yields changes in the relative height by a factor of $H/R\sim 0.1$. The main frequency of these oscillations is on a timescale of $\sim 5-6~\Omega^{-1}$ (about an orbital period), the Large R simulation shows some higher frequency behaviour at about $\sim 2-3~\Omega^{-1}$, but the dominate frequency is still around a timescale of $\sim 5~\Omega^{-1}$, which we confirm using a Fourier transform of the height time series. 

\begin{figure}
    \centering
    \includegraphics[width=\columnwidth]{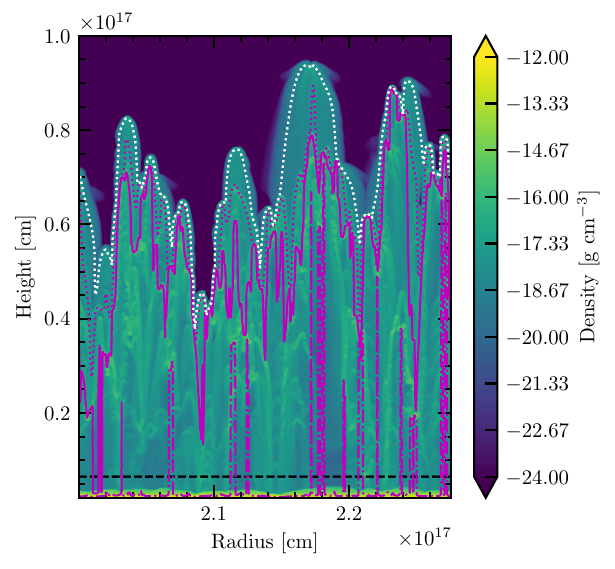}
    \caption{A density snapshot of the Large R simulation after 30 dynamical times. The black dashed line shows the position of the photosphere to the disc's thermal emission for the static initial profile. The magenta dotted, solid and dash-dotted lines shown the positions of the $\tau=0.1$, $1$ and $10$ to the disc's emission. The white dotted line shows the photosphere to radiation from the central source. Thus, despite the disc's atmosphere being initially optically thin, it evolves to a marginally optically thick case.}
    \label{fig:tau_IR}
\end{figure}

Finally, in Figure~\ref{fig:tau_IR} we overlay lines of constant optical depth on the density profile for our Large R simulation after 30 dynamical timescales. Thus, despite the disc's atmosphere being initially optically thin, in the dynamically evolved state it becomes marginally optically thick. This arises because in the dynamic state the compressed clumps have high densities, and hence high dust-to-gas ratios, yielding higher extinctions than the initial state. Although this invalidates our optically thin assumption (see ~\ref{sec:simplifications}), which should be addressed in future simulations, it potentially has important consequences for our instability. Raising the optical depth will likely change the temperature structure. Photons will escape through lower density regions \citep{Witt1996}, resulting in lower mean intensities in the higher density regions and hence lower temperatures. These lower temperatures would result in enhanced dust condensation, higher opacities, and larger accelerations, potentially increasing the velocity dispersion of the clumps. 

Furthermore, the position of the photosphere to the high-energy radiation from the central region of the AGN shows dramatic variations in height (recall that in our simulation, we ray-trace in the vertical direction). This structure will clearly lead to shadowing, which we cannot capture in our approach and could also have important dynamical consequences. 

\section{Discussion}\label{sec:discuss}
Dust condensation in the atmospheres of AGN has been suggested as the origin of the broad line region \citep[e.g.][]{Czerny2011}, where the increase in opacity from complete dust condensation yields a dramatic increase in radiation pressure that can exceed the local strength of gravity. Radiation pressure was hypothesised to launch ``clouds'' from the surface of the disc. When these clumps are exposed to radiation from the central source, the dust sublimates, switching off their radiative acceleration. Modelling \citep[e.g.][]{Naddaf2021,Naddaf2022,Naddaf2024} showed that the clumps generally fail to reach escape velocity. Only in a small ‘wind launching region’ at high Eddington accretion rates can they escape \citep{Naddaf2021}. Most clumps instead receive enough radiation to sublimate their dust and then fall back towards the disc, potentially creating observable signatures \citep[e.g.][]{Muller2022}. Although this suggestion is appealing and has some observation support, given that the broad-line region appears at distances at which dust would be expected to condense \citep[e.g.][]{Czerny2011,Czerny2015,Naddaf2020}, how exactly this process could happen dynamically and how the clumps could be generated. 

In this work, we have shown that dust condensation depends on both density and temperature in an AGN atmosphere. As a result, condensation does not occur at a single position in the atmosphere. Instead, it develops over an extended region (10–100 pressure scale heights), where the dust fraction gradually increases with altitude and decreasing density. This produces an atmosphere in which variable dust condensation can perfectly balance vertical gravity.

However, such an extended radiation pressure supported atmosphere is unstable to compressive (density) perturbations. Because the thermal timescale is short, the atmosphere is effectively locally isothermal. In this case, increasing the gas density accelerates dust condensation (since the rate scales roughly linearly with density), while the sublimation rate remains fixed, being sensitive only to temperature. Thus, higher gas densities produce higher dust-to-gas ratios, higher opacities, and a radiation pressure force that exceeds gravity locally, resulting in acceleration of the gas.  

 Our hydrodynamic simulations demonstrate this linear instability.  In the non-linear state, turbulent ``fountains'' of dusty material are accelerated to $\sim 1000$ km s$^{-1}$, with significant RMS turbulent velocities. Once exposed to the central source, this material is ionized and falls back onto the disc. Therefore, our new linear instability provides a mechanism for generating clumpy media, which our simulations qualitatively show behaves like the ``failed wind'' scenario of the broad-line region. In particular, our simulations naturally produce a turbulent environment, with high velocities similar to those that have been inserted into phenomenological modelling to explain line profiles \citep{Wu2025}. One implication of the failed wind scenario is that heavy elements produced by local star formation \citep[e.g.][]{Goodman2003,AliDib2023} are not transported to large radii by an escaping wind. This causes the metallicity of the broad-line region to become enhanced, while the metallicity of distant locations (e.g. in the vicinity of the narrow line region) remains lower. The broad-line region lies at radii similar to those where star formation is expected to occur. However, it does not transport metal-enriched gas to large radii through a wind. This is consistent with the observed metallicity differences between the broad-line and narrow-line regions \citep{Huang2023}.

Thus, while our simulations and instability provide a promising physical grounding for the suggestion that the broad-line region is the failed wind suggested by \citet{Czerny2011}, limitations of our simulations mean that we cannot confirm that our scenario is the explanation for the broad-line region. 


\subsection{The role of the central radiation}\label{sec:central_radiation_discuss}
\citet{Baskin2018} explored the role the nature of the central source had on the height above the disc where dust would be fully sublimated, showing that either a disc-like or isotropic illumination had an order-unity impact on the point at which sublimation occurred. In this work, we have introduced an additional sensitivity. Since the (unobscured) flux from the central source can vastly exceed that locally emitted from the accretion disc, some of this flux can be absorbed by the disc's surface and re-radiated to lower heights. This situation is similar to the very inner regions of protoplanetary discs, where the star's radiation controls the temperature at the disc's surface; however, the high optical depth to the disc's mid-plane ensures its still actively heated by accretion \citep[e.g.][]{Chrenko2024}, leading to a temperature inversion in the upper layers \citep{Jankovic2021}.  The amount of the central source's flux the disc's surface absorbs is sensitive to the ``flaring angle'' of the disc, and we have shown that changes to the flaring angle can lead to dramatic changes in the height of the disc's surface in our static models. Given that the flaring angle depends on the global disc structure, the time dependence we have observed in our models will likely lead to a complex interplay between the central source's luminosity (and any time dependence) and the structure of the disc's atmosphere at distances where dust is able to condense.

Our simulations indicate that the disc height oscillates with an amplitude of $z/R \sim 0.1$ on timescales comparable to the local orbital period. Depending on whether orbital phases align or not, this can cause the flaring angle to increase, decrease, or even produce shadowing, all on similar orbital timescales. Now, while the dynamical timescale to change the density structure in response to changing illumination is of the order $\Omega^{-1}$, the thermal timescale is rapid, along with the dust particle sublimation timescale (which is effectively instantaneous, \citealt{Baskin2018}). Thus, the thermal and extinction properties of the atmospheric region can respond rapidly to changes in illumination from the central source. 

Furthermore, we note that our simulations of the fountain-like structures in the evolving atmospheres show they are optically thick to the central source radiation. These fountain-like structures are radially narrow and tall (i.e. $z/\Delta R \gg 1$). In contrast to a smooth disc, which has a small flaring angle ($\ll 1$), these fountain-like structures are tall enough to be fully illuminated by the central source (with a flaring angle $\sim 1$). Their large height allows them to cast long shadows. The interplay of one fountain shadowing another, and the resulting secondary accelerations, produces highly complex, non-linear global evolution.

Our static and dynamic simulations show an increasing height of the disc's surface with radius. In fact, our static structure is similar to that found by \citet{Baskin2018}, given the similar physics at play. However, we have found that the region between the disc's photosphere to its own radiation and the point at which dust sublimates is a region of gradual dust condensation, rather than occurring a fixed location. \citet{Starkey2023} used both reverberation mapping and SED fitting simultaneously to determine that NGC 5548 possesses a highly flared or rippled structure near the dust sublimation radius. This is exactly what our model produces: a highly flared disc structure, on top of which are the fountains, creating a ripple-like structure. Further reverberation mapping and SED fitting based on the \citet{Starkey2023} model, using our simulations as a guide to set the flaring structure and ripple structure, will help determine whether this model can accurately describe the structure of a large number of AGN. 

\subsection{Implications for the broad-line region}\label{sec:implications_blr}
Our results have demonstrated that the dusty failed wind scenario of \citet{Czerny2011} could be the physical mechanism that gives rise to the broad-line region. In particular, we generate an extended clumpy region above the disc's photosphere with a large velocity dispersion powered by radiation pressure on the dusty gas. One limitation of our simulations is that the clumps only reach densities of $10^{8}$–$10^{9}$ cm$^{-3}$. These values are significantly lower than the $\sim 10^{11}$ cm$^{-3}$ inferred from AGN observations \citep{Adhikari2016}. As discussed in Appendix~\ref{appendix:resolve}, due to computational limitations, we do not believe that our simulations have fully resolved the density perturbations. 

Furthermore, \citet{Baskin2018} demonstrated that the densities of ionized clouds in the broad-line region could be reached by radiative compression due to the central source, a physical process not captured in our simulations. Thus, what we can do is verify that our simulated clumps contain sufficient mass. Our Large R simulation has a resolution of $\Delta R = \Delta z \sim 7\times10^{13}~$cm, implying a minimum mass in our simulated clumps of $\sim 5\times 10^{26}~g$ (assuming that the clumps are restricted by the grid scale to have a volume of $8\Delta z^3$, since \texttt{qcon}=2). Taking a spherical cloud with a density of $10^{11}$~cm$^{-3}$ with this minimum mass would imply a cloud with a column density of $9\times10^{23}$~cm$^{-2}$. Given that the observed clouds contain a column density of $\sim 10^{23}~$cm$^{-2}$ \citet{Bianchi2012}, it is clear our simulated clumps have sufficient mass to explain the observed ones. 

The broad-line region has a typical covering fraction of $\sim 0.3$ \citep[e.g.][]{Maiolino2001}. Given our simulations from Inner R to Large R span the range of expected radii for the launching of the failed wind in this scenario, the fact that the height of the clumpy region reaches 0.1-0.2 at small radii and 0.4-0.5 at larger radii, then our simulations are at the correct level to match the expected covering fraction, especially as individual ``fountains'' in our simulation are optically thick to the radiation from the central source. 

A notable feature of our simulations is the long-timescale variability of the disc atmosphere’s height. This variability occurs on the local dynamical timescale, typically of order years. These local variations could lead to shadowed regions of the disc atmosphere becoming exposed to the central source on years-long timescales. Given that the thermal and dust simulation timescale is fast, previously shadowed, clumpy dusty material could rapidly become ionized broad-line region clouds when exposed to the central source. 

\subsection{Variability}

The AGN unification model \citep[e.g.][]{Antonucci1993,Urry1995,Netzer2015} has had great success in explaining the link between Type 1 and Type 2 AGN in optical and unobscured / obscured AGN in X-rays in terms of an obscuring dusty torus \citep[e.g.][]{Stalevski2016}. However, recently it has become clear that some AGN can undergo dramatic changes in their optical / UV and X-ray emission properties on short timescales \citep[e.g.][]{Matt2003,Amrutha2024}. These ``changing-look'' AGNs imply that the unification model is incomplete. Changing obscuration AGNs show large changes in column density to the X-ray source \citep[e.g.][]{Matt2003}, while changing-state AGNs show a link between the central source's emission and the presence or absence of the broad line region \citep[e.g.][]{Graham2020}. 

Because the gas structure responds (and varies intrinsically, Figure~\ref{fig:height_time}) on the local dynamical timescale ($\gtrsim$ years), variations on shorter timescales cannot be caused by dynamical flows in the gas distribution. Given that many changing-look events occur on timescales shorter than the local dynamical timescale ($\sim$ months--years) \citep[e.g.][]{Ricci2023}, then they cannot be associated with large-scale structural changes in the gas distribution in our failed wind. 

The fact that the gas distribution cannot change does not mean that its properties and interaction with radiation are unable to rapidly vary. For example, when exposed to an increase in radiation from the central source, the dust in the gas can sublimate, allowing the radiation to penetrate deeper. The sublimation of dust can occur rapidly, on an essentially instantaneous time scale compared to other time scales of interest \citep{Baskin2018}. Furthermore, the gas will be heated to $\sim 10^4$~K and compressed by radiation pressure \citep{Baskin2014}. 
The flux is so high that the heating rate is effectively dominated by the timescale for the photoelectrons to collisionally thermalise, which for electrons in gas with a density of $10^{8}$~cm$^{-3}$ is a few seconds. Thus, the gas will be rapidly heated and ionized. 

The now heated gas will be compressed by radiation pressure. The compression timescale is of order:
\begin{equation}
t_{\rm comp}\sim \sqrt{\frac{Rc}{\kappa F_R}}     
\end{equation}
Adopting the Thompson scattering opacity for $\kappa$, and taking $F_R\sim 3\times 10^9~$erg~s$^{-1}$~cm$^{-2}$ (of order the minimum required to sublimate the dust) we find a compression timescale of $\sim $ months for a cloud, initially a few $10^{13}~$cm in size, similar to the light-crossing timescale of the broad line region. Thus, this compression timescale is the limiting timescale, suggesting that our initially dusty, clumpy disc atmosphere can respond, changing its ionization state and clump sizes on a timescale of order $\sim$ months to a change in the central source's luminosity. This timescale is similar to the response time of some changing-look AGNs. Thus, below we speculate how our scenario might lead to changing obscuration and changing-state AGN.  

\begin{figure*}
    \centering
    \includegraphics[width=\textwidth, trim={0 27.7cm 19.2cm 0},clip]{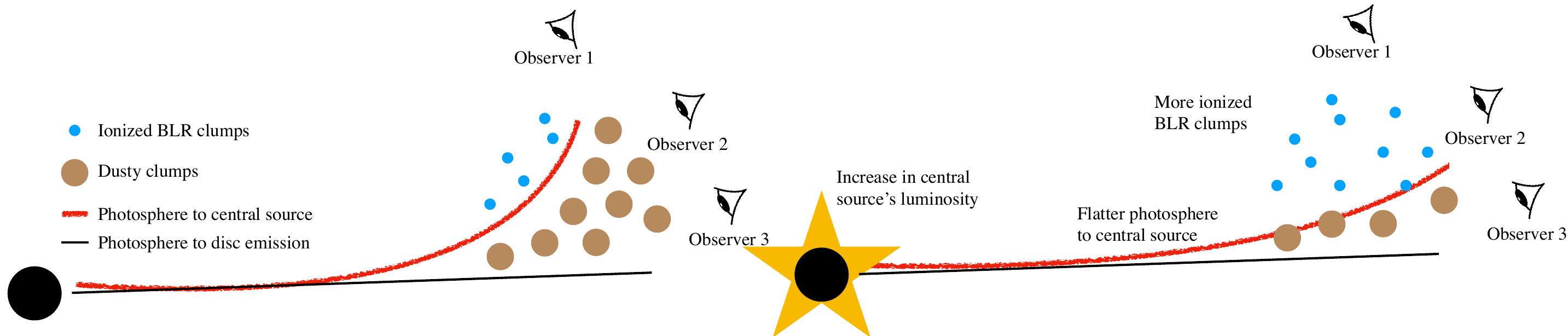}
    \caption{Schematic concept of our AGN disc atmosphere structure response to a rapid increase in the central source's luminosity and the generation of a changing-look AGN. The increase in the central source's luminosity sublimates dust in the disc's atmosphere, pushing its photosphere closer to the disc's photosphere to its own radiation. Due to the rapid change, the clumps do not move in response to the increase in luminosity; however, now exposed clumps become ionized and radiatively compressed. Observer 1 would see an increase in the BLR emission in response to the increase in central source's luminosity. Observer 2 would observe a changing state AGN (Type 2 $\rightarrow$ Type 1), along with a bluer optical continuum due to the flatter disc photosphere to the central source. Finally, Observer 3 may see a changing obscuration AGN as the radiatively compressed clumps would have a lower covering fraction than the larger, previously dusty clumps. Observers 2 and 3 would observe changing obscuration AGN without any change in the central source's luminosity as ionized or dusty clumps rotate into and out of their line of sight.  }\label{fig:cartoon}
\end{figure*}

Therefore, qualitatively at least, the dusty, clumpy failed wind we have identified will respond to changes in the central source's radiation in a manner similar to that of a changing-look AGN. A sketch of the response of the disc atmosphere, as observed, to changes in the central source's luminosity is shown in Figure~\ref{fig:cartoon}, which we detail below.  

\subsubsection{Change in obscuration due to the orbital motion of clumps}

Since the size of the X-ray source can be quite small $\lesssim 10^{13}$~cm \citep[e.g.][]{Risaliti2009}, many short-timescale changing obscuration AGN are thought to be linked to clouds moving in and out of the line of sight \citep[e.g.][]{Ricci2023}. Since our disc atmosphere is clumpy for essentially all of its height (e.g. Figure~\ref{fig:tau_IR}), with clump sizes of $\sim 10^{13}-10^{14}$~cm and column densities at the $10^{21}-10^{23}$~cm$^{-2}$ level, then our clumpy disc atmosphere could certainly provide an explanation for changing-look AGN with changes in column density of order $\Delta N\sim10^{22}-10^{23}$~cm$^{-2}$, where just clouds moving in and out of line-of-sight are responsible for the rapid changes \citep[e.g.][]{Risalit2005,Risaliti2007}. However, it seems unlikely that this can fully explain some of the large changes in column-density $\Delta N\gtrsim 10^{24}$~cm$^{-2}$ observed.

\subsubsection{Change in obscuration due to change in the central source's luminosity}

Some AGN show very large changes in the obscuring column ($\Delta N\gtrsim 10^{24}$~cm$^{-2}$) which might be associated with a change in emission from the central source \citep[e.g.][]{Guainazzi1998,Gilli2000}. 

For changing obscuration AGN associated with a change in the central source's luminosity,  the required change in the central source's ionising radiation is often insufficient to explain the observed change in column depth by ionizing the material \citep{Risalit2005}. Although large changes in the obscuring column depth can occur due to the large scale changes in the amplitude of the disc's height identified in our simulations (e.g. Figure~\ref{fig:height_time}), this can happen only on order of the local dynamical timescale ($\sim $ years). Alternatively, we can speculate that during a change in the central source's luminosity which sublimates the dust in the atmosphere (rapidly), there is a change in covering fraction associated with the compression of the previously lower density dusty clumps into the dense ionized clumps that form the broad-line region clouds.  

If we assume in the dusty region that the clumps have an internal density of $n_{d}$ with size $r_{d}$, then with a number density of clumps $n_c$, the effective extinction coefficient through the clumpy media is \citep{Hobson1993}:
\begin{equation}
\chi_{d,c} = n_c\pi r_d^2 f_{\rm int}   
\end{equation}
with,
\begin{equation}
    f_{\rm int} = 1 + \frac{\exp\left(-2\tau_{c}\right)}{\tau_{\rm c}}-\frac{1-\exp\left(-2\tau_{c}\right)}{2\tau_c^2} 
\end{equation}
where $\tau_{c}$ is the surface to centre extinction optical depth for the clump, under the assumption of spherical clumps. In the case of the dusty clumps, $\tau_c=\tau_{c,d}=\sigma_d n_d r_d$, where $\sigma_d$ is the extinction cross-section of the dusty gas. 

\begin{figure}
    \centering
    \includegraphics[width=\columnwidth]{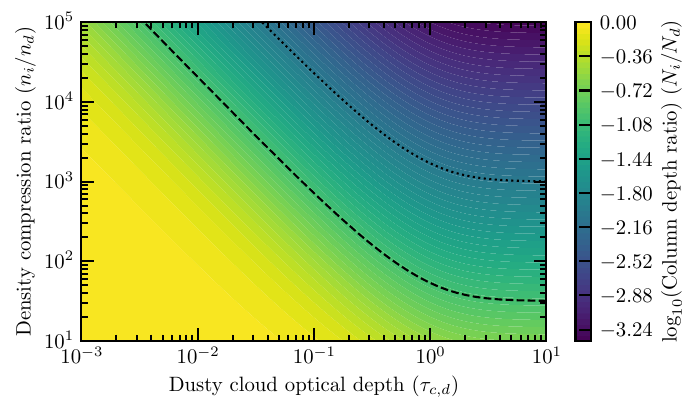}
    \caption{The change in the obscuring column density in the X-rays due to radiative compression during dust sublimation of the clumpy outflow. Calculated under the assumption the clouds do not fragment ($M=1$) and the extinction coefficient of the medium does not change ($\sigma_i/\sigma_d=1$). The dashed line shows a drop in the column density of a factor of 10, while the dashed line shows a drop of 100. Thus, radiative compression of the dusty clumps by ionizing radiation can cause a large change in the medium's effective extinction column density, without needing to change the gas opacity.}
    \label{fig:column_density_change}
\end{figure}

Any increase in the central source's UV/optical emission should result in it sublimating the dust in the clumps and the ionizing radiation field being exposed to the previous clumpy, dusty gas causing it to become dust free. As discussed above, these previously dusty clumps, when exposed to the central source, can undergo radiative compression to much higher densities, causing them to shrink considerably, resulting in a lower effective extinction coefficient. If we suppose that each dusty clump is broken into $M$ ionized clumps, during radiative compression, then the effective extinction coefficient through the now ionized clumpy media (with clumps with internal densities $n_i$ and sizes $r_i$), becomes:
\begin{equation}
    \chi_{i,c}=Mn_c\pi r_i^2 f_{\rm int}\left(\tau_{c,i}\right) = M^{1/3}n_c\pi \left(\frac{n_d}{n_i}\right)^{2/3} r_d^2 f_{\rm int}\left(\tau_{c,i}\right)
\end{equation}
Therefore, the change in column density due purely to the compression of the clumps is:
\begin{equation}
    \frac{N_i}{N_d} = M^{1/3} \left(\frac{n_d}{n_i}\right)^{2/3} \frac{f_{\rm int}\left(\tau_{c,i}\right)}{f_{\rm int}\left(\tau_{c,d}\right)}
\end{equation}
which, yields a large change in column-depth in the limit the individual clumps become optically thick ($f(\tau)\rightarrow 1$), whereas, as expected in the optically thin limit the change in column density scales with the change in extinction coefficient going from dusty to ionized gas. We plot the relative ratio of column density for different density ratios in Figure~\ref{fig:column_density_change}, without clump fragmentation and without change in gas opacity due to ionization. 

Our typical dusty clump densities and sizes from our simulations yields an individual clump column density of $\sim 10^{21}-10^{22}$~cm$^{-2}$, which has an optical depth of $\sim 0.1-1$ at 1~keV. Figure~\ref{fig:column_density_change} shows that the large compression ratios expected between dusty and ionized clumps ($\gtrsim 10^4$) can produce large changes in the effective column density. 

\subsubsection{Change in the optical}

Changing state AGN often show the appearance of the broad emission lines and the optical continuum becoming bluer in response to an increase in the central source's luminosity \citep[e.g.][]{MacLeod2016}. Such a response could be a natural outcome of the response of our disc's atmosphere to an increase in the central source's luminosity. 

As discussed above, an increase in the central source's optical/UV output will sublimate dust in the disc's atmosphere, allowing it to penetrate deeper. This has multiple effects: (i) significantly reduces the optical column density to an inclined observer (relative to the disc's axis of symmetry); (ii) ionizes and radiatively compresses the previous dusty clumps generated low-ionization broad-line emission clumps over a significantly larger volume; and (iii) reduces the flaring angle of the disc's photosphere to the central source's radiation. 

Thus, the impact of the first two results will significantly increase the number of broad-line emitting clouds in the observers' line-of-sight, where the increase will be larger for more inclined observers. Furthermore, the reduction in the flaring angle evolution with radius ($\mu\propto R^n$) will cause less of the central source's radiation to be absorbed, and as such reprocessed at larger distances; hence the flatter disc surface (smaller $n$) will give rise to a bluer reprocessing spectrum.

\subsection{Simplifications and future directions}\label{sec:simplifications}

Although we have identified a new instability in the atmosphere of AGN due to dust condensation, that would give rise to a failed wind as envisaged by \citet{Czerny2011}; whether or not our method can provide the physical explanation for the broad-line region, and associated AGN phenomena like changing-look AGN remains open. In order to investigate this problem, both analytically and numerically, we have made a number of simplifying assumptions that are worth exploring.

Currently, our opacity model is linked to a parametrised dust condensation-sublimation model, where the dust fraction is a smooth function of temperature and density. The exact structure of our static atmosphere shows a weak dependence with the sharpness of this smoothing. While the sublimation and condensation timescales are so fast that a vapour pressure equilibrium can be thought to hold, the exact smoothness of the transition will depend on the details of the size distribution and opacity (which itself depends on the size distribution). The densities in our atmosphere are not too dissimilar to those in protoplanetary disc mid-planes where we know the size distribution is capable of evolving dynamically through coagulation and fragmentation \citep[e.g.][]{Birnstiel2024}. Thus, future calculations should consider how the dust size distribution might evolve dynamically and how this feeds back on both the sublimation-condensation equilibrium and the opacities. 

Furthermore, we have assumed that the dust and gas are fully dynamically coupled. Since radiation pressure primarily acts of the dust particles, which then transmit this acceleration to the gas through collisions, there is a coupling time scale between the dust and gas. If this coupling is too slow, dust can become partially decoupled from the gas, moving with a different velocity, which in turn could trigger secondary streaming instabilities \citep[e.g.][]{Hopkins2018,Squire2018}. Small, micron-sized particles have large Knudsen numbers such that the particles are in the Epstein drag limit, where the stopping time is:
\begin{equation}
    t_{\rm stop} = \frac{a \rho_{\rm in}}{v_{\rm th}\rho}\sim 0.7~{\rm yrs} \left(\frac{a}{1~{\rm \mu m}}\right) \left(\frac{n}{10^7~{\rm cm}^{-3}}\right)^{-1}\left(\frac{T}{1500~{\rm K}}\right)^{-1/2}
\end{equation}
where $a$ is the particle size, $\rho_{\rm in}$ is the internal density of the dust particles (set to 2~g~cm$^{-3}$), and $v_{\rm th}$ is the gas' mean thermal velocity. This timescale is short enough that there should be dynamical coupling between gas and dust, especially in the denser clumps where the densities $\gtrsim 10^{8}$~cm$^{-3}$, in the lower density regions, or if the dust particles were able to grow sufficiently large then the gas and dust might start to become decoupled. Thus, future analytic and simulation work should explore the role of imperfect coupling between the dust and the gas. Such decoupling might lead to even higher contrasts in the dust density, potentially enhancing the clumpiness and acceleration of the high density regions.  

In our analytic work and simulations we neglected the background shear in the disc for simplicity. However, this is not too bad an assumption. Shear acts on of order the gas pressure scale height of the disc, which while a few percent of the radius, is still much larger than the scales of the instability we have found (whose fastest growing mode is at $kH\sim 100$, Figure~\ref{fig:numerical}). Furthermore, because the clumps reach speeds in excess of 1000~km~s$^{-1}$, we should also consider the role of the Coriolis force. The Coriolis turning-length:
\begin{equation}
    R_c = \frac{|\mathbf{u}|}{2\Omega}
\end{equation}
is even larger than the pressure scale height (by a factor of $|\mathbf{u}|/2c_s$). Thus, while we do not expect neglecting the disc's background shear and the disc's rotation to fundamentally change the nature of our instability or results, future simulation work should explore the resulting impact on the non-linear state of our failed-wind, as a proper treatment of the full geometry of the wind is critical to matching the observations \citep[e.g.][]{Naddaf2021}.

One of the biggest drawbacks of our results is the local geometry and resolutions of our simulations. The central source's radiation is implicitly included through vertical ray-tracing at a fixed flaring angle. However, the clumpy nature of our disc surface means that one must dynamically evolve the flaring angle, a result that could only be achieved through a more global simulation approach. However, any global simulation would worsen our resolution. Even in our local simulation boxes, to simulate down to the scale of the radiation compress ionized clumps would require a dynamic range approaching or exceeding a factor of $10^4$. Thus, exploring the full range of physics involved, from shadowing to radiative compression of the sublimated dusty clumps will likely require expensive (and probably hierarchical) simulations, while extending our 2D approach to 3D. 

\section{Conclusions} \label{sec:summary}

We have studied dust condensation in the atmospheres of AGN discs. By including a density-dependent treatment of dust condensation, motivated by physical condensation--sublimation models, we have shown that the atmospheres of AGN discs are large, extended structures in which dust condensation slowly occurs, increasing with height. By analysing the stability of these atmospheres and studying their dynamics in 2D hydrodynamic simulations, we have shown that they are unstable to a radiation-pressure-induced dust condensation instability. In the non-linear state, the disc's atmosphere represents a clumpy ``failed wind'' similar in properties and dynamics to that proposed to explain the low-ionization broad-line region \citep[e.g.][]{Czerny2011}. The main findings of our work are:
\begin{itemize}
    \item In static models of the vertical structure of a disc, once the effective temperature approaches the dust condensation threshold ($\sim 1,500$~K), dust begins to condense. However, as the appearance of a small amount of dust dramatically increases the opacity, the atmosphere becomes radiation-pressure supported, with vertical gravity essentially balanced by radiation pressure. As the height in the disc increases, the dust abundance increases to balance the stronger gravity. This results in large, extended disc atmospheres (over 10s of pressure scale heights) where the density and temperature slowly decrease while the dust abundance increases. 

    \item This dusty disc atmosphere is linearly unstable. Perturbations that increase (decrease) the density raise (lower) the dust sublimation temperature, leading to larger (smaller) dust abundances. As higher dust abundance increases the opacity, a density increase can cause the perturbation to be accelerated vertically. 

    \item This compressive instability operates when the perturbation is close to isothermal and has a growth rate that scales with wavenumber as $\sqrt{k}$. Given the rapid thermal timescales in the disc's atmosphere, the isothermal assumption is valid down to extremely small wavelengths ($kH\sim 100$). 

    \item 2D hydrodynamic simulations confirm the presence of this linear instability. In the non-linear state, the disc's atmosphere evolves into a highly clumpy and dynamic configuration where radiation pressure can accelerate ``clouds'' to velocities of $\sim 1000$~km~s$^{-1}$ before the dust is sublimated and the material falls back towards the disc.

    \item The surface of the disc's atmosphere evolves on the local dynamical timescale, with its height changing by a factor of $\sim 2$. In addition, fountains of dusty material are launched above the disc surface on smaller scales. These phenomena create a highly flared, rippled, and dynamic disc surface that is likely to cast strongly time-variable shadows further out on the disc. 

    \item We have speculated that this clumpy disc atmosphere could lead to the low-ionization broad-line region and potentially explain some of the changing-look phenomena recently observed in AGN populations. 
\end{itemize}

While we are confident that the instability we have identified is robust and likely to occur in AGN discs, determining whether it naturally explains the observed broad-line region, changing-look AGN, and other phenomena requires further work with more detailed, higher-resolution simulations.

\section*{Acknowledgements}
We are grateful to the referee for comments that improved the manuscript. JEO is supported by a Royal Society University Research Fellowship. We thank Keith Horne for
useful conversations.  This work benefited from the 2023 Exoplanet Summer Program in the Other Worlds Laboratory (OWL) at the University of California, Santa Cruz, a program funded by the Heising-Simons Foundation.

\section*{Data Availability}

 The data underlying this article will be shared with reasonable request to the corresponding author.



\bibliographystyle{mnras}
\bibliography{example} 

\begin{thebibliography}{}
\makeatletter
\relax
\def\mn@urlcharsother{\let\do\@makeother \do\$\do\&\do\#\do\^\do\_\do\%\do\~}
\def\mn@doi{\begingroup\mn@urlcharsother \@ifnextchar [ {\mn@doi@} {\mn@doi@[]}}
\def\mn@doi@[#1]#2{\def\@tempa{#1}\ifx\@tempa\@empty \href {http://dx.doi.org/#2} {doi:#2}\else \href {http://dx.doi.org/#2} {#1}\fi \endgroup}
\def\mn@eprint#1#2{\mn@eprint@#1:#2::\@nil}
\def\mn@eprint@arXiv#1{\href {http://arxiv.org/abs/#1} {{\tt arXiv:#1}}}
\def\mn@eprint@dblp#1{\href {http://dblp.uni-trier.de/rec/bibtex/#1.xml} {dblp:#1}}
\def\mn@eprint@#1:#2:#3:#4\@nil{\def\@tempa {#1}\def\@tempb {#2}\def\@tempc {#3}\ifx \@tempc \@empty \let \@tempc \@tempb \let \@tempb \@tempa \fi \ifx \@tempb \@empty \def\@tempb {arXiv}\fi \@ifundefined {mn@eprint@\@tempb}{\@tempb:\@tempc}{\expandafter \expandafter \csname mn@eprint@\@tempb\endcsname \expandafter{\@tempc}}}

\bibitem[\protect\citeauthoryear{{Adhikari}, {R{\'o}{\.z}a{\'n}ska}, {Czerny}, {Hryniewicz}  \& {Ferland}}{{Adhikari} et~al.}{2016}]{Adhikari2016}
{Adhikari} T.~P.,  {R{\'o}{\.z}a{\'n}ska} A.,  {Czerny} B.,  {Hryniewicz} K.,   {Ferland} G.~J.,  2016, \mn@doi [\apj] {10.3847/0004-637X/831/1/68}, \href {https://ui.adsabs.harvard.edu/abs/2016ApJ...831...68A} {831, 68}

\bibitem[\protect\citeauthoryear{{Ali}}{{Ali}}{2021}]{Ali2021}
{Ali} A.~A.,  2021, \mn@doi [\mnras] {10.1093/mnras/staa3992}, \href {https://ui.adsabs.harvard.edu/abs/2021MNRAS.501.4136A} {501, 4136}

\bibitem[\protect\citeauthoryear{{Ali-Dib} \& {Lin}}{{Ali-Dib} \& {Lin}}{2023}]{AliDib2023}
{Ali-Dib} M.,  {Lin} D. N.~C.,  2023, \mn@doi [\mnras] {10.1093/mnras/stad2774}, \href {https://ui.adsabs.harvard.edu/abs/2023MNRAS.526.5824A} {526, 5824}

\bibitem[\protect\citeauthoryear{{Amrutha}, {Wolf}, {Onken}, {Hon}, {Lai}, {Tonry}  \& {Webster}}{{Amrutha} et~al.}{2024}]{Amrutha2024}
{Amrutha} N.,  {Wolf} C.,  {Onken} C.~A.,  {Hon} W.~J.,  {Lai} S.,  {Tonry} J.~L.,   {Webster} R.,  2024, \mn@doi [\mnras] {10.1093/mnras/stae2470}, \href {https://ui.adsabs.harvard.edu/abs/2024MNRAS.535.2322A} {535, 2322}

\bibitem[\protect\citeauthoryear{{Antonucci}}{{Antonucci}}{1993}]{Antonucci1993}
{Antonucci} R.,  1993, \mn@doi [\araa] {10.1146/annurev.aa.31.090193.002353}, \href {https://ui.adsabs.harvard.edu/abs/1993ARA&A..31..473A} {31, 473}

\bibitem[\protect\citeauthoryear{{Baskin} \& {Laor}}{{Baskin} \& {Laor}}{2018}]{Baskin2018}
{Baskin} A.,  {Laor} A.,  2018, \mn@doi [\mnras] {10.1093/mnras/stx2850}, \href {https://ui.adsabs.harvard.edu/abs/2018MNRAS.474.1970B} {474, 1970}

\bibitem[\protect\citeauthoryear{{Baskin}, {Laor}  \& {Stern}}{{Baskin} et~al.}{2014}]{Baskin2014}
{Baskin} A.,  {Laor} A.,   {Stern} J.,  2014, \mn@doi [\mnras] {10.1093/mnras/stt2230}, \href {https://ui.adsabs.harvard.edu/abs/2014MNRAS.438..604B} {438, 604}

\bibitem[\protect\citeauthoryear{{Begelman}, {McKee}  \& {Shields}}{{Begelman} et~al.}{1983}]{Begelman1983}
{Begelman} M.~C.,  {McKee} C.~F.,   {Shields} G.~A.,  1983, \mn@doi [\apj] {10.1086/161178}, \href {https://ui.adsabs.harvard.edu/abs/1983ApJ...271...70B} {271, 70}

\bibitem[\protect\citeauthoryear{{Bell} \& {Lin}}{{Bell} \& {Lin}}{1994}]{Bell1994}
{Bell} K.~R.,  {Lin} D.~N.~C.,  1994, \mn@doi [\apj] {10.1086/174206}, \href {https://ui.adsabs.harvard.edu/abs/1994ApJ...427..987B} {427, 987}

\bibitem[\protect\citeauthoryear{{Bianchi}, {Maiolino}  \& {Risaliti}}{{Bianchi} et~al.}{2012}]{Bianchi2012}
{Bianchi} S.,  {Maiolino} R.,   {Risaliti} G.,  2012, \mn@doi [Advances in Astronomy] {10.1155/2012/782030}, \href {https://ui.adsabs.harvard.edu/abs/2012AdAst2012E..17B} {2012, 782030}

\bibitem[\protect\citeauthoryear{{Birnstiel}}{{Birnstiel}}{2024}]{Birnstiel2024}
{Birnstiel} T.,  2024, \mn@doi [\araa] {10.1146/annurev-astro-071221-052705}, \href {https://ui.adsabs.harvard.edu/abs/2024ARA&A..62..157B} {62, 157}

\bibitem[\protect\citeauthoryear{{Blandford} \& {Payne}}{{Blandford} \& {Payne}}{1982}]{Blandford1982}
{Blandford} R.~D.,  {Payne} D.~G.,  1982, \mn@doi [\mnras] {10.1093/mnras/199.4.883}, \href {https://ui.adsabs.harvard.edu/abs/1982MNRAS.199..883B} {199, 883}

\bibitem[\protect\citeauthoryear{{Campos Estrada}, {Owen}, {Jankovic}, {Wilson}  \& {Helling}}{{Campos Estrada} et~al.}{2024}]{Campos2024}
{Campos Estrada} B.,  {Owen} J.~E.,  {Jankovic} M.~R.,  {Wilson} A.,   {Helling} C.,  2024, \mn@doi [\mnras] {10.1093/mnras/stae095}, \href {https://ui.adsabs.harvard.edu/abs/2024MNRAS.528.1249C} {528, 1249}

\bibitem[\protect\citeauthoryear{{Chiang} \& {Goldreich}}{{Chiang} \& {Goldreich}}{1997}]{Chiang1997}
{Chiang} E.~I.,  {Goldreich} P.,  1997, \mn@doi [\apj] {10.1086/304869}, \href {https://ui.adsabs.harvard.edu/abs/1997ApJ...490..368C} {490, 368}

\bibitem[\protect\citeauthoryear{{Chiang}, {Joung}, {Creech-Eakman}, {Qi}, {Kessler}, {Blake}  \& {van Dishoeck}}{{Chiang} et~al.}{2001}]{Chiang2001}
{Chiang} E.~I.,  {Joung} M.~K.,  {Creech-Eakman} M.~J.,  {Qi} C.,  {Kessler} J.~E.,  {Blake} G.~A.,   {van Dishoeck} E.~F.,  2001, \mn@doi [\apj] {10.1086/318427}, \href {https://ui.adsabs.harvard.edu/abs/2001ApJ...547.1077C} {547, 1077}

\bibitem[\protect\citeauthoryear{{Chrenko}, {Flock}, {Ueda}, {M{\'e}rand}, {Benisty}  \& {Chametla}}{{Chrenko} et~al.}{2024}]{Chrenko2024}
{Chrenko} O.,  {Flock} M.,  {Ueda} T.,  {M{\'e}rand} A.,  {Benisty} M.,   {Chametla} R.~O.,  2024, \mn@doi [\aj] {10.3847/1538-3881/ad234d}, \href {https://ui.adsabs.harvard.edu/abs/2024AJ....167..124C} {167, 124}

\bibitem[\protect\citeauthoryear{{Collin} \& {Zahn}}{{Collin} \& {Zahn}}{2008}]{Collin2008}
{Collin} S.,  {Zahn} J.~P.,  2008, \mn@doi [\aap] {10.1051/0004-6361:20078191}, \href {https://ui.adsabs.harvard.edu/abs/2008A&A...477..419C} {477, 419}

\bibitem[\protect\citeauthoryear{{Collin-Souffrin}, {Dyson}, {McDowell}  \& {Perry}}{{Collin-Souffrin} et~al.}{1988}]{CS1988}
{Collin-Souffrin} S.,  {Dyson} J.~E.,  {McDowell} J.~C.,   {Perry} J.~J.,  1988, \mn@doi [\mnras] {10.1093/mnras/232.3.539}, \href {https://ui.adsabs.harvard.edu/abs/1988MNRAS.232..539C} {232, 539}

\bibitem[\protect\citeauthoryear{{Czerny} \& {Hryniewicz}}{{Czerny} \& {Hryniewicz}}{2011}]{Czerny2011}
{Czerny} B.,  {Hryniewicz} K.,  2011, \mn@doi [\aap] {10.1051/0004-6361/201016025}, \href {https://ui.adsabs.harvard.edu/abs/2011A&A...525L...8C} {525, L8}

\bibitem[\protect\citeauthoryear{{Czerny} et~al.,}{{Czerny} et~al.}{2015}]{Czerny2015}
{Czerny} B.,  et~al., 2015, \mn@doi [Advances in Space Research] {10.1016/j.asr.2015.01.004}, \href {https://ui.adsabs.harvard.edu/abs/2015AdSpR..55.1806C} {55, 1806}

\bibitem[\protect\citeauthoryear{{Czerny}, {Du}, {Wang}  \& {Karas}}{{Czerny} et~al.}{2016}]{Czerny2016}
{Czerny} B.,  {Du} P.,  {Wang} J.-M.,   {Karas} V.,  2016, \mn@doi [\apj] {10.3847/0004-637X/832/1/15}, \href {https://ui.adsabs.harvard.edu/abs/2016ApJ...832...15C} {832, 15}

\bibitem[\protect\citeauthoryear{{Dobbs-Dixon}, {Agol}  \& {Burrows}}{{Dobbs-Dixon} et~al.}{2012}]{DobbsDixon2012}
{Dobbs-Dixon} I.,  {Agol} E.,   {Burrows} A.,  2012, \mn@doi [\apj] {10.1088/0004-637X/751/2/87}, \href {https://ui.adsabs.harvard.edu/abs/2012ApJ...751...87D} {751, 87}

\bibitem[\protect\citeauthoryear{{Done} \& {Krolik}}{{Done} \& {Krolik}}{1996}]{Done1996}
{Done} C.,  {Krolik} J.~H.,  1996, \mn@doi [\apj] {10.1086/177230}, \href {https://ui.adsabs.harvard.edu/abs/1996ApJ...463..144D} {463, 144}

\bibitem[\protect\citeauthoryear{{Elvis}, {Marengo}  \& {Karovska}}{{Elvis} et~al.}{2002}]{Elvis2002}
{Elvis} M.,  {Marengo} M.,   {Karovska} M.,  2002, \mn@doi [\apjl] {10.1086/340006}, \href {https://ui.adsabs.harvard.edu/abs/2002ApJ...567L.107E} {567, L107}

\bibitem[\protect\citeauthoryear{{Gaskell} \& {Goosmann}}{{Gaskell} \& {Goosmann}}{2016}]{Gaskell2016}
{Gaskell} C.~M.,  {Goosmann} R.~W.,  2016, \mn@doi [\apss] {10.1007/s10509-015-2648-1}, \href {https://ui.adsabs.harvard.edu/abs/2016Ap&SS.361...67G} {361, 67}

\bibitem[\protect\citeauthoryear{{Gilli}, {Maiolino}, {Marconi}, {Risaliti}, {Dadina}, {Weaver}  \& {Colbert}}{{Gilli} et~al.}{2000}]{Gilli2000}
{Gilli} R.,  {Maiolino} R.,  {Marconi} A.,  {Risaliti} G.,  {Dadina} M.,  {Weaver} K.~A.,   {Colbert} E.~J.~M.,  2000, \mn@doi [\aap] {10.48550/arXiv.astro-ph/0001107}, \href {https://ui.adsabs.harvard.edu/abs/2000A&A...355..485G} {355, 485}

\bibitem[\protect\citeauthoryear{{Goodman}}{{Goodman}}{2003}]{Goodman2003}
{Goodman} J.,  2003, \mn@doi [\mnras] {10.1046/j.1365-8711.2003.06241.x}, \href {https://ui.adsabs.harvard.edu/abs/2003MNRAS.339..937G} {339, 937}

\bibitem[\protect\citeauthoryear{{Graham} et~al.,}{{Graham} et~al.}{2020}]{Graham2020}
{Graham} M.~J.,  et~al., 2020, \mn@doi [\mnras] {10.1093/mnras/stz3244}, \href {https://ui.adsabs.harvard.edu/abs/2020MNRAS.491.4925G} {491, 4925}

\bibitem[\protect\citeauthoryear{{Gravity Collaboration} et~al.,}{{Gravity Collaboration} et~al.}{2018}]{GravityAGN2018}
{Gravity Collaboration} et~al., 2018, \mn@doi [\nat] {10.1038/s41586-018-0731-9}, \href {https://ui.adsabs.harvard.edu/abs/2018Natur.563..657G} {563, 657}

\bibitem[\protect\citeauthoryear{{Gravity Collaboration} et~al.,}{{Gravity Collaboration} et~al.}{2021}]{GravityAGN2021}
{Gravity Collaboration} et~al., 2021, \mn@doi [\aap] {10.1051/0004-6361/202040061}, \href {https://ui.adsabs.harvard.edu/abs/2021A&A...648A.117G} {648, A117}

\bibitem[\protect\citeauthoryear{{Grier} et~al.,}{{Grier} et~al.}{2013}]{Grier2013}
{Grier} C.~J.,  et~al., 2013, \mn@doi [\apj] {10.1088/0004-637X/764/1/47}, \href {https://ui.adsabs.harvard.edu/abs/2013ApJ...764...47G} {764, 47}

\bibitem[\protect\citeauthoryear{{Guainazzi} et~al.,}{{Guainazzi} et~al.}{1998}]{Guainazzi1998}
{Guainazzi} M.,  et~al., 1998, \mn@doi [\mnras] {10.1046/j.1365-8711.1998.02089.x}, \href {https://ui.adsabs.harvard.edu/abs/1998MNRAS.301L...1G} {301, L1}

\bibitem[\protect\citeauthoryear{{Hayes}, {Norman}, {Fiedler}, {Bordner}, {Li}, {Clark}, {ud-Doula}  \& {Mac Low}}{{Hayes} et~al.}{2006}]{Hayes2006}
{Hayes} J.~C.,  {Norman} M.~L.,  {Fiedler} R.~A.,  {Bordner} J.~O.,  {Li} P.~S.,  {Clark} S.~E.,  {ud-Doula} A.,   {Mac Low} M.-M.,  2006, \mn@doi [\apjs] {10.1086/504594}, \href {https://ui.adsabs.harvard.edu/abs/2006ApJS..165..188H} {165, 188}

\bibitem[\protect\citeauthoryear{{Hobson} \& {Padman}}{{Hobson} \& {Padman}}{1993}]{Hobson1993}
{Hobson} M.~P.,  {Padman} R.,  1993, \mn@doi [\mnras] {10.1093/mnras/264.1.161}, \href {https://ui.adsabs.harvard.edu/abs/1993MNRAS.264..161H} {264, 161}

\bibitem[\protect\citeauthoryear{{Hopkins} \& {Squire}}{{Hopkins} \& {Squire}}{2018}]{Hopkins2018}
{Hopkins} P.~F.,  {Squire} J.,  2018, \mn@doi [\mnras] {10.1093/mnras/sty1982}, \href {https://ui.adsabs.harvard.edu/abs/2018MNRAS.480.2813H} {480, 2813}

\bibitem[\protect\citeauthoryear{{Huang}, {Lin}  \& {Shields}}{{Huang} et~al.}{2023}]{Huang2023}
{Huang} J.,  {Lin} D. N.~C.,   {Shields} G.,  2023, \mn@doi [\mnras] {10.1093/mnras/stad2642}, \href {https://ui.adsabs.harvard.edu/abs/2023MNRAS.525.5702H} {525, 5702}

\bibitem[\protect\citeauthoryear{{Hubeny}}{{Hubeny}}{1990}]{Hubeny1990}
{Hubeny} I.,  1990, \mn@doi [\apj] {10.1086/168501}, \href {https://ui.adsabs.harvard.edu/abs/1990ApJ...351..632H} {351, 632}

\bibitem[\protect\citeauthoryear{{Isella} \& {Natta}}{{Isella} \& {Natta}}{2005}]{Isella2005}
{Isella} A.,  {Natta} A.,  2005, \mn@doi [\aap] {10.1051/0004-6361:20052773}, \href {https://ui.adsabs.harvard.edu/abs/2005A&A...438..899I} {438, 899}

\bibitem[\protect\citeauthoryear{{Jankovic}, {Owen}, {Mohanty}  \& {Tan}}{{Jankovic} et~al.}{2021}]{Jankovic2021}
{Jankovic} M.~R.,  {Owen} J.~E.,  {Mohanty} S.,   {Tan} J.~C.,  2021, \mn@doi [\mnras] {10.1093/mnras/stab920}, \href {https://ui.adsabs.harvard.edu/abs/2021MNRAS.504..280J} {504, 280}

\bibitem[\protect\citeauthoryear{{Jankovic}, {Mohanty}, {Owen}  \& {Tan}}{{Jankovic} et~al.}{2022}]{Jankovic2022}
{Jankovic} M.~R.,  {Mohanty} S.,  {Owen} J.~E.,   {Tan} J.~C.,  2022, \mn@doi [\mnras] {10.1093/mnras/stab3370}, \href {https://ui.adsabs.harvard.edu/abs/2022MNRAS.509.5974J} {509, 5974}

\bibitem[\protect\citeauthoryear{{K{\"a}ppeli} \& {Mishra}}{{K{\"a}ppeli} \& {Mishra}}{2014}]{Kappeli2014}
{K{\"a}ppeli} R.,  {Mishra} S.,  2014, \mn@doi [Journal of Computational Physics] {10.1016/j.jcp.2013.11.028}, \href {https://ui.adsabs.harvard.edu/abs/2014JCoPh.259..199K} {259, 199}

\bibitem[\protect\citeauthoryear{{K{\"a}ppeli} \& {Mishra}}{{K{\"a}ppeli} \& {Mishra}}{2016}]{Kappeli2016}
{K{\"a}ppeli} R.,  {Mishra} S.,  2016, \mn@doi [\aap] {10.1051/0004-6361/201527815}, \href {https://ui.adsabs.harvard.edu/abs/2016A&A...587A..94K} {587, A94}

\bibitem[\protect\citeauthoryear{{Kaspi}, {Maoz}, {Netzer}, {Peterson}, {Vestergaard}  \& {Jannuzi}}{{Kaspi} et~al.}{2005}]{Kaspi2005}
{Kaspi} S.,  {Maoz} D.,  {Netzer} H.,  {Peterson} B.~M.,  {Vestergaard} M.,   {Jannuzi} B.~T.,  2005, \mn@doi [\apj] {10.1086/431275}, \href {https://ui.adsabs.harvard.edu/abs/2005ApJ...629...61K} {629, 61}

\bibitem[\protect\citeauthoryear{{Kollatschny}}{{Kollatschny}}{2003}]{Kollatschny2003}
{Kollatschny} W.,  2003, \mn@doi [\aap] {10.1051/0004-6361:20030928}, \href {https://ui.adsabs.harvard.edu/abs/2003A&A...407..461K} {407, 461}

\bibitem[\protect\citeauthoryear{{Korista}, {Baldwin}, {Ferland}  \& {Verner}}{{Korista} et~al.}{1997}]{Korista1997}
{Korista} K.,  {Baldwin} J.,  {Ferland} G.,   {Verner} D.,  1997, \mn@doi [\apjs] {10.1086/312966}, \href {https://ui.adsabs.harvard.edu/abs/1997ApJS..108..401K} {108, 401}

\bibitem[\protect\citeauthoryear{{Kuiper}, {Klahr}, {Beuther}  \& {Henning}}{{Kuiper} et~al.}{2010}]{Kuiper2010}
{Kuiper} R.,  {Klahr} H.,  {Beuther} H.,   {Henning} T.,  2010, \mn@doi [\apj] {10.1088/0004-637X/722/2/1556}, \href {https://ui.adsabs.harvard.edu/abs/2010ApJ...722.1556K} {722, 1556}

\bibitem[\protect\citeauthoryear{{MacLeod} et~al.,}{{MacLeod} et~al.}{2016}]{MacLeod2016}
{MacLeod} C.~L.,  et~al., 2016, \mn@doi [\mnras] {10.1093/mnras/stv2997}, \href {https://ui.adsabs.harvard.edu/abs/2016MNRAS.457..389M} {457, 389}

\bibitem[\protect\citeauthoryear{{Maiolino}, {Salvati}, {Marconi}  \& {Antonucci}}{{Maiolino} et~al.}{2001}]{Maiolino2001}
{Maiolino} R.,  {Salvati} M.,  {Marconi} A.,   {Antonucci} R.~R.~J.,  2001, \mn@doi [\aap] {10.1051/0004-6361:20010808}, \href {https://ui.adsabs.harvard.edu/abs/2001A&A...375...25M} {375, 25}

\bibitem[\protect\citeauthoryear{{Maiolino} et~al.,}{{Maiolino} et~al.}{2010}]{Maiolino2010}
{Maiolino} R.,  et~al., 2010, \mn@doi [\aap] {10.1051/0004-6361/200913985}, \href {https://ui.adsabs.harvard.edu/abs/2010A&A...517A..47M} {517, A47}

\bibitem[\protect\citeauthoryear{{Matt}, {Guainazzi}  \& {Maiolino}}{{Matt} et~al.}{2003}]{Matt2003}
{Matt} G.,  {Guainazzi} M.,   {Maiolino} R.,  2003, \mn@doi [\mnras] {10.1046/j.1365-8711.2003.06539.x}, \href {https://ui.adsabs.harvard.edu/abs/2003MNRAS.342..422M} {342, 422}

\bibitem[\protect\citeauthoryear{{Mignone}, {Zanni}, {Tzeferacos}, {van Straalen}, {Colella}  \& {Bodo}}{{Mignone} et~al.}{2012}]{pluto_code}
{Mignone} A.,  {Zanni} C.,  {Tzeferacos} P.,  {van Straalen} B.,  {Colella} P.,   {Bodo} G.,  2012, \mn@doi [\apjs] {10.1088/0067-0049/198/1/7}, \href {https://ui.adsabs.harvard.edu/abs/2012ApJS..198....7M} {198, 7}

\bibitem[\protect\citeauthoryear{{M{\"u}ller}, {Naddaf}, {Zaja{\v{c}}ek}, {Czerny}, {Araudo}  \& {Karas}}{{M{\"u}ller} et~al.}{2022}]{Muller2022}
{M{\"u}ller} A.~L.,  {Naddaf} M.-H.,  {Zaja{\v{c}}ek} M.,  {Czerny} B.,  {Araudo} A.,   {Karas} V.,  2022, \mn@doi [\apj] {10.3847/1538-4357/ac660a}, \href {https://ui.adsabs.harvard.edu/abs/2022ApJ...931...39M} {931, 39}

\bibitem[\protect\citeauthoryear{{Murray}, {Chiang}, {Grossman}  \& {Voit}}{{Murray} et~al.}{1995}]{Murray1995}
{Murray} N.,  {Chiang} J.,  {Grossman} S.~A.,   {Voit} G.~M.,  1995, \mn@doi [\apj] {10.1086/176238}, \href {https://ui.adsabs.harvard.edu/abs/1995ApJ...451..498M} {451, 498}

\bibitem[\protect\citeauthoryear{{Naddaf}}{{Naddaf}}{2024}]{Naddaf2024}
{Naddaf} M.~H.,  2024, \mn@doi [arXiv e-prints] {10.48550/arXiv.2412.18772}, \href {https://ui.adsabs.harvard.edu/abs/2024arXiv241218772N} {p. arXiv:2412.18772}

\bibitem[\protect\citeauthoryear{{Naddaf} \& {Czerny}}{{Naddaf} \& {Czerny}}{2022}]{Naddaf2022}
{Naddaf} M.~H.,  {Czerny} B.,  2022, \mn@doi [\aap] {10.1051/0004-6361/202142806}, \href {https://ui.adsabs.harvard.edu/abs/2022A&A...663A..77N} {663, A77}

\bibitem[\protect\citeauthoryear{{Naddaf} \& {Czerny}}{{Naddaf} \& {Czerny}}{2024}]{NC2024}
{Naddaf} M.-H.,  {Czerny} B.,  2024, \mn@doi [Universe] {10.3390/universe10010029}, \href {https://ui.adsabs.harvard.edu/abs/2024Univ...10...29N} {10, 29}

\bibitem[\protect\citeauthoryear{{Naddaf}, {Czerny}  \& {Szczerba}}{{Naddaf} et~al.}{2020}]{Naddaf2020}
{Naddaf} M.-H.,  {Czerny} B.,   {Szczerba} R.,  2020, \mn@doi [Frontiers in Astronomy and Space Sciences] {10.3389/fspas.2020.00015}, \href {https://ui.adsabs.harvard.edu/abs/2020FrASS...7...15N} {7, 15}

\bibitem[\protect\citeauthoryear{{Naddaf}, {Czerny}  \& {Szczerba}}{{Naddaf} et~al.}{2021}]{Naddaf2021}
{Naddaf} M.-H.,  {Czerny} B.,   {Szczerba} R.,  2021, \mn@doi [\apj] {10.3847/1538-4357/ac139d}, \href {https://ui.adsabs.harvard.edu/abs/2021ApJ...920...30N} {920, 30}

\bibitem[\protect\citeauthoryear{{Netzer}}{{Netzer}}{1990}]{Netzer1990}
{Netzer} H.,  1990, in {Blandford} R.~D.,  {Netzer} H.,  {Woltjer} L.,  {Courvoisier} T.~J.~L.,   {Mayor} M.,  eds, Active Galactic Nuclei. pp 57--160

\bibitem[\protect\citeauthoryear{{Netzer}}{{Netzer}}{2015}]{Netzer2015}
{Netzer} H.,  2015, \mn@doi [\araa] {10.1146/annurev-astro-082214-122302}, \href {https://ui.adsabs.harvard.edu/abs/2015ARA&A..53..365N} {53, 365}

\bibitem[\protect\citeauthoryear{{Netzer} \& {Laor}}{{Netzer} \& {Laor}}{1993}]{Netzer1993}
{Netzer} H.,  {Laor} A.,  1993, \mn@doi [\apjl] {10.1086/186741}, \href {https://ui.adsabs.harvard.edu/abs/1993ApJ...404L..51N} {404, L51}

\bibitem[\protect\citeauthoryear{{Owen}}{{Owen}}{2020}]{Owen2020_snow}
{Owen} J.~E.,  2020, \mn@doi [\mnras] {10.1093/mnras/staa1309}, \href {https://ui.adsabs.harvard.edu/abs/2020MNRAS.495.3160O} {495, 3160}

\bibitem[\protect\citeauthoryear{{Padovani} et~al.,}{{Padovani} et~al.}{2017}]{Padovani2017}
{Padovani} P.,  et~al., 2017, \mn@doi [\aapr] {10.1007/s00159-017-0102-9}, \href {https://ui.adsabs.harvard.edu/abs/2017A&ARv..25....2P} {25, 2}

\bibitem[\protect\citeauthoryear{{Panda}, {Czerny}, {Adhikari}, {Hryniewicz}, {Wildy}, {Kuraszkiewicz}  \& {{\'S}niegowska}}{{Panda} et~al.}{2018}]{Panda2018}
{Panda} S.,  {Czerny} B.,  {Adhikari} T.~P.,  {Hryniewicz} K.,  {Wildy} C.,  {Kuraszkiewicz} J.,   {{\'S}niegowska} M.,  2018, \mn@doi [\apj] {10.3847/1538-4357/aae209}, \href {https://ui.adsabs.harvard.edu/abs/2018ApJ...866..115P} {866, 115}

\bibitem[\protect\citeauthoryear{{Peterson}}{{Peterson}}{2006}]{Peterson2006}
{Peterson} B.~M.,  2006, in {Alloin} D.,  ed., , Vol.~693, Physics of Active Galactic Nuclei at all Scales.
p.~77, \mn@doi{10.1007/3-540-34621-X_3}

\bibitem[\protect\citeauthoryear{{Peterson} et~al.,}{{Peterson} et~al.}{2004}]{Peterson2004}
{Peterson} B.~M.,  et~al., 2004, \mn@doi [\apj] {10.1086/423269}, \href {https://ui.adsabs.harvard.edu/abs/2004ApJ...613..682P} {613, 682}

\bibitem[\protect\citeauthoryear{{Ricci} \& {Trakhtenbrot}}{{Ricci} \& {Trakhtenbrot}}{2023}]{Ricci2023}
{Ricci} C.,  {Trakhtenbrot} B.,  2023, \mn@doi [Nature Astronomy] {10.1038/s41550-023-02108-4}, \href {https://ui.adsabs.harvard.edu/abs/2023NatAs...7.1282R} {7, 1282}

\bibitem[\protect\citeauthoryear{{Risaliti}, {Elvis}, {Fabbiano}, {Baldi}  \& {Zezas}}{{Risaliti} et~al.}{2005}]{Risalit2005}
{Risaliti} G.,  {Elvis} M.,  {Fabbiano} G.,  {Baldi} A.,   {Zezas} A.,  2005, \mn@doi [\apjl] {10.1086/430252}, \href {https://ui.adsabs.harvard.edu/abs/2005ApJ...623L..93R} {623, L93}

\bibitem[\protect\citeauthoryear{{Risaliti}, {Elvis}, {Fabbiano}, {Baldi}, {Zezas}  \& {Salvati}}{{Risaliti} et~al.}{2007}]{Risaliti2007}
{Risaliti} G.,  {Elvis} M.,  {Fabbiano} G.,  {Baldi} A.,  {Zezas} A.,   {Salvati} M.,  2007, \mn@doi [\apjl] {10.1086/517884}, \href {https://ui.adsabs.harvard.edu/abs/2007ApJ...659L.111R} {659, L111}

\bibitem[\protect\citeauthoryear{{Risaliti} et~al.,}{{Risaliti} et~al.}{2009}]{Risaliti2009}
{Risaliti} G.,  et~al., 2009, \mn@doi [\mnras] {10.1111/j.1745-3933.2008.00580.x}, \href {https://ui.adsabs.harvard.edu/abs/2009MNRAS.393L...1R} {393, L1}

\bibitem[\protect\citeauthoryear{{Ruff}, {Floyd}, {Webster}, {Korista}  \& {Landt}}{{Ruff} et~al.}{2012}]{Ruff2012}
{Ruff} A.~J.,  {Floyd} D. J.~E.,  {Webster} R.~L.,  {Korista} K.~T.,   {Landt} H.,  2012, \mn@doi [\apj] {10.1088/0004-637X/754/1/18}, \href {https://ui.adsabs.harvard.edu/abs/2012ApJ...754...18R} {754, 18}

\bibitem[\protect\citeauthoryear{{Schulik} \& {Booth}}{{Schulik} \& {Booth}}{2023}]{Schulik2023}
{Schulik} M.,  {Booth} R.~A.,  2023, \mn@doi [\mnras] {10.1093/mnras/stad1251}, \href {https://ui.adsabs.harvard.edu/abs/2023MNRAS.523..286S} {523, 286}

\bibitem[\protect\citeauthoryear{{Shlosman} \& {Begelman}}{{Shlosman} \& {Begelman}}{1989}]{Shlosman1989}
{Shlosman} I.,  {Begelman} M.~C.,  1989, \mn@doi [\apj] {10.1086/167526}, \href {https://ui.adsabs.harvard.edu/abs/1989ApJ...341..685S} {341, 685}

\bibitem[\protect\citeauthoryear{{Squire} \& {Hopkins}}{{Squire} \& {Hopkins}}{2018}]{Squire2018}
{Squire} J.,  {Hopkins} P.~F.,  2018, \mn@doi [\apjl] {10.3847/2041-8213/aab54d}, \href {https://ui.adsabs.harvard.edu/abs/2018ApJ...856L..15S} {856, L15}

\bibitem[\protect\citeauthoryear{{Stalevski}, {Ricci}, {Ueda}, {Lira}, {Fritz}  \& {Baes}}{{Stalevski} et~al.}{2016}]{Stalevski2016}
{Stalevski} M.,  {Ricci} C.,  {Ueda} Y.,  {Lira} P.,  {Fritz} J.,   {Baes} M.,  2016, \mn@doi [\mnras] {10.1093/mnras/stw444}, \href {https://ui.adsabs.harvard.edu/abs/2016MNRAS.458.2288S} {458, 2288}

\bibitem[\protect\citeauthoryear{{Starkey}, {Huang}, {Horne}  \& {Lin}}{{Starkey} et~al.}{2023}]{Starkey2023}
{Starkey} D.~A.,  {Huang} J.,  {Horne} K.,   {Lin} D. N.~C.,  2023, \mn@doi [\mnras] {10.1093/mnras/stac3579}, \href {https://ui.adsabs.harvard.edu/abs/2023MNRAS.519.2754S} {519, 2754}

\bibitem[\protect\citeauthoryear{{Stone} \& {Norman}}{{Stone} \& {Norman}}{1992}]{Stone1992}
{Stone} J.~M.,  {Norman} M.~L.,  1992, \mn@doi [\apjs] {10.1086/191680}, \href {https://ui.adsabs.harvard.edu/abs/1992ApJS...80..753S} {80, 753}

\bibitem[\protect\citeauthoryear{{Stone}, {Tomida}, {White}  \& {Felker}}{{Stone} et~al.}{2020}]{athena_code}
{Stone} J.~M.,  {Tomida} K.,  {White} C.~J.,   {Felker} K.~G.,  2020, \mn@doi [\apjs] {10.3847/1538-4365/ab929b}, \href {https://ui.adsabs.harvard.edu/abs/2020ApJS..249....4S} {249, 4}

\bibitem[\protect\citeauthoryear{{Suganuma} et~al.,}{{Suganuma} et~al.}{2006}]{Suganuma2006}
{Suganuma} M.,  et~al., 2006, \mn@doi [\apj] {10.1086/499326}, \href {https://ui.adsabs.harvard.edu/abs/2006ApJ...639...46S} {639, 46}

\bibitem[\protect\citeauthoryear{{Sulentic}, {Marziani}  \& {Dultzin-Hacyan}}{{Sulentic} et~al.}{2000}]{Sulentic2000}
{Sulentic} J.~W.,  {Marziani} P.,   {Dultzin-Hacyan} D.,  2000, \mn@doi [\araa] {10.1146/annurev.astro.38.1.521}, \href {https://ui.adsabs.harvard.edu/abs/2000ARA&A..38..521S} {38, 521}

\bibitem[\protect\citeauthoryear{{Urry} \& {Padovani}}{{Urry} \& {Padovani}}{1995}]{Urry1995}
{Urry} C.~M.,  {Padovani} P.,  1995, \mn@doi [\pasp] {10.1086/133630}, \href {https://ui.adsabs.harvard.edu/abs/1995PASP..107..803U} {107, 803}

\bibitem[\protect\citeauthoryear{{Wang}, {Du}, {Baldwin}, {Ge}, {Hu}  \& {Ferland}}{{Wang} et~al.}{2012}]{Wang2012}
{Wang} J.-M.,  {Du} P.,  {Baldwin} J.~A.,  {Ge} J.-Q.,  {Hu} C.,   {Ferland} G.~J.,  2012, \mn@doi [\apj] {10.1088/0004-637X/746/2/137}, \href {https://ui.adsabs.harvard.edu/abs/2012ApJ...746..137W} {746, 137}

\bibitem[\protect\citeauthoryear{{Wang}, {Du}, {Brotherton}, {Hu}, {Songsheng}, {Li}, {Shi}  \& {Zhang}}{{Wang} et~al.}{2017}]{Wang2017}
{Wang} J.-M.,  {Du} P.,  {Brotherton} M.~S.,  {Hu} C.,  {Songsheng} Y.-Y.,  {Li} Y.-R.,  {Shi} Y.,   {Zhang} Z.-X.,  2017, \mn@doi [Nature Astronomy] {10.1038/s41550-017-0264-4}, \href {https://ui.adsabs.harvard.edu/abs/2017NatAs...1..775W} {1, 775}

\bibitem[\protect\citeauthoryear{{Witt} \& {Gordon}}{{Witt} \& {Gordon}}{1996}]{Witt1996}
{Witt} A.~N.,  {Gordon} K.~D.,  1996, \mn@doi [\apj] {10.1086/177282}, \href {https://ui.adsabs.harvard.edu/abs/1996ApJ...463..681W} {463, 681}

\bibitem[\protect\citeauthoryear{{Wu}, {Wu}, {Lu}, {Cao}, {Lei}, {Wang}  \& {Fan}}{{Wu} et~al.}{2025}]{Wu2025}
{Wu} J.,  {Wu} Q.,  {Lu} K.-X.,  {Cao} X.,  {Lei} X.,  {Wang} M.,   {Fan} X.,  2025, \mn@doi [\apj] {10.3847/1538-4357/ada271}, \href {https://ui.adsabs.harvard.edu/abs/2025ApJ...979..125W} {979, 125}

\makeatother
\end{thebibliography}



\appendix
\section{Resolution of the simulations}\label{appendix:resolve}

\begin{figure}
    \centering
    \includegraphics[width=\columnwidth]{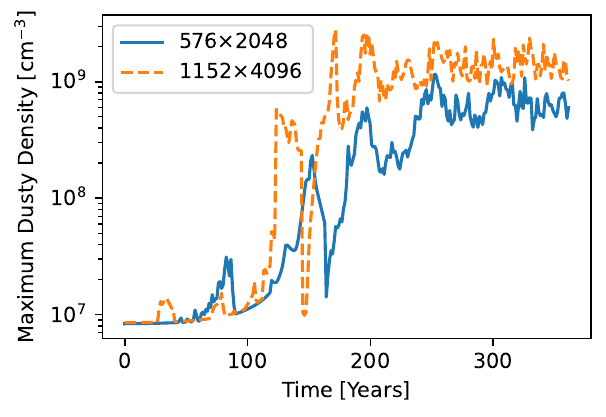}
    \caption{The time evolution of the maximum density in our Large R simulation with the standard resolution (solid) and with double the resolution (dashed). The higher-resolution case evolves faster, by resolving faster-growing modes of the instability, and reaches higher densities. }
    \label{fig:resolution}
\end{figure}

If we were to include full radiative transfer in our simulation, allowing us to fully resolve the fastest-growing mode ($kH\sim 100$), this would require a cell size of $\lesssim 10^{12}~$cm in the Large R simulation. Given the physical requirement to include a height of order $10^{17}$~cm in the simulation box, this would necessitate $\sim 10^5$ cells along a coordinate direction. Since the majority of the domain is fully turbulent, an adaptive-mesh-refinement approach would be of limited benefit. Thus, we must assess which aspects of our simulations are converged and which are not. We re-ran the Large R simulation with twice the resolution of the original (i.e. $N_R\times N_z=1152\times4096$). The behaviour of the simulation is the same—we see a linear growth phase (albeit with a faster growth rate due to the smaller scales now resolved) and a non-linear phase of fountains of dusty material launched from the disc atmosphere. We find the RMS velocity dispersion is identical, and the temporal evolution of the disc height oscillates with a similar amplitude and timescale, indicating these effects are likely resolved. However, the density in the fountains is larger in the higher-resolution simulations. Essentially, radiation pressure is compressing the clumps towards the grid scale. In Figure~\ref{fig:resolution}, we show the evolution of the maximum density in the dusty fountains as a function of time for the two simulations. 

The fact that the same velocity dispersion is found is expected, as the clumps have densities sufficiently high in both simulations for dust to fully condense, and as such the radiative acceleration is identical. Given that they essentially evolve ballistically under radiative and gravitational acceleration, we find the same velocities, which is reassuring for interpreting the kinematics of our failed wind. 

However, the densities are about a factor of $\sim 2$ higher in the higher-resolution simulation. Given that these simulations are 2D, the cell volume has shrunk by a factor of 4 when the resolution was doubled. Thus, the clumps are less massive in the higher-resolution simulations (by a factor of $\sim 2$). Therefore, at this stage it is unclear what the densities and masses of the clumps in our failed wind and dusty fountains will attain in reality. This extremely high-resolution requirement will make determining this challenging, especially since this should also be explored with 3D simulations. It is likely that extremely local simulations will need to be run. 

\bsp	
\label{lastpage}
\end{document}